\documentclass[11pt,a4paper]{article}

\usepackage{array}  
\usepackage{acronym}
\usepackage{algorithm}  
\usepackage{algorithmic}
\usepackage{amsmath}           
\usepackage{amssymb}		
\usepackage[titletoc]{appendix}
\usepackage{a4wide}
 \usepackage{array}             
 \usepackage{caption}  
 \usepackage[]{hyperref}  
 \usepackage{graphicx}
\usepackage{natbib}
\usepackage{lscape}
\usepackage{subfigure}  
\usepackage{times}
\usepackage[flushleft]{threeparttable}
\usepackage{verbatim}

\newcounter{rowno}
\graphicspath{{./images/}}

 \DeclareGraphicsExtensions{.pdf,.jpeg,.png}

\newcounter{lastnote}

\title{Using Twitter to Model the EUR/USD Exchange Rate}

\author
{Dietmar Janetzko\\
\\
\normalsize{Cologne Business School, Cologne, Germany}\\
\normalsize{dietmarja@gmail.com}
\thanks{I like to thank Bertold Albrecht, Sarah Conn, Bipin Indurkhya, Julia Maintz and Stephan Weibelzahl  for insightful comments on a previous version of this paper. }
}

\begin{document}

\newcommand{\NumberofConcepts}{594}
\newcommand{\NumberofFinalConcepts}{17}
\newcommand{\DaysofDataRecording}{636}
\newcommand{\BeginofDataRecording}{January 1, 2012}
\newcommand{\EndofDataRecording}{September 27, 2013}
\renewcommand*{\bflabel}[1]{{#1\hfill}}   
\renewcommand*\appendixpagename{Appendix} 
\renewcommand*\appendixtocname{Appendix}
\newcommand*\mean[1]{\bar{#1}}
\newcommand{\mylinespacing}{5}
\newcommand{\argmax}[1]{\underset{#1}{\operatorname{arg}\,\operatorname{max}}\;}

\maketitle
\begin{abstract}
Fast, global, and sensitively reacting  to political, economic and social events of any kind, these are attributes that social media like Twitter share with foreign exchange markets.
Does the former allow us to predict the latter? 
The leading assumption of this paper is that time series of Tweet counts have predictive content for exchange rate movements. 
This assumption prompted a Twitter-based exchange rate model that harnesses regARIMA analyses for short-term out-of-sample ex post forecasts of the daily closing prices of  EUR/USD spot exchange rates. The analyses made use of Tweet counts collected from \BeginofDataRecording{} -- \EndofDataRecording{} 
via the Otter API of {\tt topsy.com}. 
To identify concepts mentioned on Twitter with a predictive potential the analysis followed a 2-step selection. Firstly, a heuristic qualitative analysis assembled a long list  of  \NumberofConcepts{}  concepts, e.g., {\it Merkel}, {\it Greece}, {\it Cyprus}, {\it crisis}, {\it chaos}, {\it growth}, {\it unemployment}  expected to covary with the ups and downs of the EUR/USD exchange rate. Secondly, cross-validation using window averaging  with a fixed-sized rolling  origin 
was deployed to select concepts and corresponding univariate time series that had error scores below chance level as defined by the random walk model that is based only on the EUR/USD exchange rate. 
With regard to a short list of 17 concepts (covariates),  in particular {\it SP} ({\it Standard \& Poor's}) and {\it risk}, the out-of-sample predictive accuracy of the Twitter-based regARIMA model was found to be repeatedly better than that obtained from both the random walk model and a random noise covariate in 1-step ahead forecasts of the EUR/USD exchange rate. This advantage was evident on 
the level of forecast error metrics (MSFE, MAE)
 when a majority vote over different estimation windows was conducted. 
The results challenge the semi-strong form of the efficient market hypothesis 
\citep{Fama_70, Fama_91} which when applied to the FX market maintains that all publicly available information is already integrated into exchange rates. 
\end{abstract}

\maketitle

\section{Introduction}
\label{Introduction}
The foreign exchange market is known to be the largest financial market in the world. Over the last decades, 
``the FX trading volume has exploded reflecting an electronic revolution that has lowered trading costs, attracted new groups of market participants, and enabled aggressive new trading strategies''
\citep[][p.\ 3 ]{King+_11}. 
In this market, the EUR/USD exchange rate\footnote{The EUR/USD spot exchange rate is stated as U.S.\ Dollars per Euro.}  is currently considered to be the most important currency pair. 
When  wars, disasters, economic decline, high unemployment rates, elections in major countries are looming, the EUR/USD exchange rate often reacts like a global crisis barometer. 
Usually, these reactions like exchange rate movements in general  can be well explained in hindsight, but over short-time periods they are hard to predict above chance level.  
In a recent comprehensive review of the vast, but not always satisfying work on exchange rate modeling Barbara Rossi underscores that success in this area depends to a large degree on the choice of predictors  \citep{Rossi_13}. 
The work described in this paper has been sparked by the question whether such predictors can be distilled from public discussions on the microblogging platform Twitter.  
Information harvested from Twitter has already become a promising  source of  data for predictive modeling of many real-world phenomena 
\citep{Lica-Tuta_11}.
Areas modeled are as diverse as forecasting box-office revenues for movies 
\citep{Asur-Huberman_10}, 
disease activity of influenza-type illnesses 
\citep{Signorini+_11}, 
stock market indicators like Dow Jones Industrial Average  
\citep{Bollen+_11}  or 
political elections 
\citep{Tumasjan+_10}. 
More closely related to the present work is a recent study by
\cite{Papaioannou_13} who used data sourced from Twitter between 25/10/2010 and 05/01/2011 to model the EUR/USD exchange on a high-frequency intraday trading scale. 
The work introduced in this paper shares the general  research goal pursued by \cite{Papaioannou_13}. It differs, however, from this study insofar as it looks at the influence of 
public debates on the EUR/USD exchange rate. Accordingly, the present study make use of a less fine grained temporal resolution (daily closing prices of  EUR/USD spot exchange rates) which are studied  
over a longer time span (21 months). The more specific goal of this paper is to address the question whether and to what extent information extracted from public debates on Twitter facilitates prediction of the 
EUR/USD exchange rate.  

 Being global and sensitive to small changes on the international news agenda Twitter  seems to fit the metaphor of a global crisis barometer.  
Can this metaphor be turned into a predictive model of the EUR/USD exchange rate?
On browsing the research on predictive modeling via data from social media  three  requirements of the domain to be modeled spring to mind: Firstly, all areas modeled should provoke {\it intensive public discussions\/} on social media platforms. Actually acquiring this data may or may not be non-trivial or costly. Clearly, however, there should  not be any shortage of data available for modeling.
Secondly, for each area to be modeled an unequivocal {\it numerical measure\/} external to the Internet, e.g., revenues, stock exchange rates or results of a general election is essential. This requirement 
emphasizes that a variable to be predicted is an indispensable part of any forecasting setup. 
Thirdly, social media usually generate an overabundance of data. Twitter, for instance, is a hub for half a billion text messages a day  
\citep{Goel_13}. The popularity of social media often means that finding information that  facilitates the modeling task chosen is  challenging. 
The third requirement for successful modeling via data from social media is  therefore the {\it selection of  information\/} with explanatory or predictive potential. For instance, \cite{Papaioannou_13} addressed the selection problem by searching via the Archivist API for Tweets that used the expression ``buy EUR/USD'' .

The three requirements discussed above help to narrow down conditions under which Twitter can be deployed to model the EUR/USD exchange rate. 
Firstly, Twitter provides a global platform where news, opinions, comments are exchanged on an unprecedented scale. 
Often, this type of information varies with the volatility of markets including the foreign exchange markets. 
Secondly, the EUR/USD exchange rate is obviously a numerical measure amenable to modeling.  
Whether or not data gleaned from Twitter can fulfill the third requirement, i.e.,  selection of  information with explanatory or predictive potential, is an open question that merits a more detailed discussion. 

The US Dollar and the Euro are mentioned in a large number of Tweets that do not seem to have any relationship with the EUR/USD exchange rate, e.g., advertisements. Vice versa, there is a large number of Tweets that do not mention the US Dollar or the Euro explicitly but which may be connected with the EUR/USD exchange rate, e.g., mentions of unemployment figures in the US or a crisis in a EU country. 
Even though the information found on Twitter is often nonsensical or simply unrelated to foreign exchange markets  it can be assumed that a sizeable number of the messages 
communicated on Twitter include information that bear some relationship with the foreign exchange markets. 
Following this assumption, the question at stake is which information exchanged on Twitter has a predictive potential with regard to variables of the foreign exchange market. 
 If this type of  information is  available on Twitter, it needs to be identified using appropriate selectors.
 What could be such an appropriate selector? 
 The research described in this paper is guided by the assumption that on Twitter the selection of variables with predictive content for the EUR/USD exchange rate  
 is achievable by referring to the concepts used in discussions of the Euro crisis. This is a specify focus, and it has to be specific to work as a selector for Tweets.  
 At the same time, this focus  does in itself not rule out that data elicited in this way may  also reflect positive developments in the Euro zone. 
 This is true as long as there are some discussions on the Euro crisis on Twitter. Otherwise, there is no variance in the data and prediction becomes impossible. 
  It is conjectured that intensive discussions of the Euro crisis are associated with a bearish Euro and a bullish US Dollar. 
 Vice versa, if the discussion on any crisis in the Euro zone is losing momentum, e.g., because of positive news from the Euro zone, then the value of the Euro relative to the US Dollar can be expected to be on the up. Furthermore, it is assumed that such discussions on the Euro crisis are not only associated with the EUR/USD exchange rate but bear potential to predict it. 

The assumption that there exists  information on Twitter  (Tweets) that facilitates predicting variables of interest  is termed here the 
hypothesis of sufficient predictive information.
This hypothesis has been inspired by the conjecture that Twitter is a fast, efficient and globally operating aggregator of news and opinions.
  Using information from Twitter to forecast  foreign exchange rates with more accuracy than the random walk model (RW) 
  is considered to be supporting evidence in favor of the hypothesis of sufficient predictive information. 
A competing approach to the prediction of foreign exchange rates that leads to opposing predictions when applied to Twitter is the well-known
{\it efficient market hypothesis\/} (EMH). 
There are three forms of the EMH, and in this paper the focus is on the semi-strong form. It maintains that investors are rational and pick up swiftly all market-relevant 
information so that the exchange rate considered fully reflects the information that is publicly available.  
According to this hypothesis, only truly new information, which occurs at random, can affect the markets. 
A corollary of the efficient market hypothesis is that the random walk model predicts market prices best. 
It is typically applied to the analysis of stock markets but is also common in the analysis of foreign exchange 
markets. However, the question of efficiency has sparked a number of controversial debates 
\citep[e.g., ][]{Lee-Sodoikhuu_12}. 

The position maintained in this paper is that the efficient market hypothesis is sufficiently general to put information gleaned from Twitter into perspective as well 
\citep[see also][]{Papaioannou_13}. 
The Twitter-based information considered in this study are Tweet counts. This type of information is not as easily accessible as information taken from, e.g., 
newspapers but it is information that is publicly available. 
The semi-strong form of the market hypothesis predicts that discussions on Twitter like any other market-relevant publicly available information is rapidly arbitraged away and impounded into exchange rates. A truly efficient foreign exchange market would  mean that discussions on Twitter have no predictive value for exchange rates. Still, harnessing Twitter for prediction would be methodologically na\"{\i}ve  without considering both the potential and the limits of the fast and dynamically changing character of discussions on this microblogging platform.  
On the one hand, data gleaned from Twitter may facilitate predictions of variables of interest. 
In fact, the very existence of Twitter is a challenge to the efficient market hypothesis as it offers real-time data with possible predictive power or predictability for foreign exchange markets and other  markets. On the other hand, the predictive power of concepts mentioned on Twitter is expected to vary across time and may easily be subject to structural breaks. 
The research question addressed in this study is whether and under what conditions the hypothesis of sufficient predictive information
that operates on the basis of data gleaned from Twitter or  the efficient market hypothesis facilitates better predictions of the EUR/USD exchange rate. The conditions to be examined include the identification of concepts that facilitate prediction of the  EUR/USD exchange rate  and the expected time-dependence  of their predictability. 
To answer this question a horse race between 
regARIMA models and the random walk model has been conducted.
The models are nested, in that the regARIMA model is an extension of the parsimonious random walk model. 
The former is taken to model the hypothesis of sufficient information, the latter models the efficient market hypothesis. 
The mean square forecast error (MSFE) and the mean absolute prediction (MAE) error of the regARIMA model relative to that of the random walk model 
are used as loss functions and criteria of forecast accuracy. 
Akaike's information criterion 
\citep[AIC,][]{Akaike_74} and the Bayesian information criterion 
\citep[BIC,][]{Schwarz_78} 
are harnessed as measures of model fit. In what follows, the expression {\it outperforming concepts} refers to concepts which when used as covariates in a regARIMA model  lead to smaller forecast error scores (MSFE, MAE) than the random walk model. Seen from this vantage point, this study is part of a strand of research in exchange rate modeling that attempts to `beat' the random walk model 
\citep{MacDonald-Taylor_94,Lisi-Medio_97,Kilian-Taylor_03,Hong+_07,Rossi_13}. At the same time, however, this study differs from this research in that the exchange rate model proposed makes use of data gathered from social media. The analysis will reveal whether or not there are outperforming concepts talked about on Twitter. 

The remainder of this article is organized around the following sections. 
To motivate the Twitter-based approach to exchange rate modeling section 2 discusses the relationship between the EUR/USD exchange rate and public discussions.
Section 3 spells out the theoretical background and the econometric models used in this study. 
In section 4, a 5-step approach to data elicitation from Twitter and feature extraction is described. 
Section 5 presents the results of this study. 
The paper concludes with a discussion that relates the findings of this study to the hypothesis of sufficient information and the efficient market hypothesis.

\section{Exchange Rates and Public Communication}
\label{Exchange_Rates_and_Public_Communication}
The existence of a link between communication and exchange rates is well known to observers of the FX market. 
 Politicians or high ranking bank officials trying to talk currencies  up or down via the news 
 is a typical example of attempts to instrumentalize this link.
 In this study, however, the focus is on Twitter-based public discussions around currencies irrespective of whether they have been prompted by politicians, bank officials or any other party. 
 Does the general public discuss the EUR/USD exchange rate? If such a discussion exists, does it influence exchange rates? Indirectly, the vast literature 
on the EUR/USD exchange rate seems to give a negative answer to this question. The majority of the academic work on exchange rates does not consider 
discussions of the general public but focuses on macro-fundamentals and to a smaller degree on non-fundamentals, e.g., news or sentiments. 
What the general public discusses has attracted far less attention among researchers of exchange rates. 
The small number of studies that do exist on public discussions  and  exchange rates examine the role of media or institutions of the finance sector on exchange rates
\citep{Thompson_09} 
or they examine tactical public communications of politicians.  But these studies do not work towards exchange rate modeling 
\citep{Bracke+_08,Weiss-Kemper_11}. 
Hence, it is an open question whether and to what degree the ups and downs of the EUR/USD are explicitly discussed by the general public. But it seems to be safe to assume that the state of the economy in the United States, in the Euro zone and beyond matters to  many people 
\citep{Knorr-Cetina-Bruegger_02}. 
 A large number of economic topics which the general public is concerned with, e.g., real income, unemployment, debts, prices, rents, wages, housing are known to be closely related to exchange rates. In fact, many of these topics directly relate to economic fundamentals. 
Until recently, there was hardly any platform available where people outside financial expert communities could articulate their thoughts and sentiments related to theses topics. The Internet, in particular the microblogging service Twitter, has changed this. Used by non-professionals and professionals alike, Twitter offers a global platform for discussing a huge variety of topics including those which directly or indirectly concern the EUR/USD exchange rate. In this study, the intensity of public debates on Twitter will be measured by the number of Tweets that include one or more of the concepts considered to be indicative for the EUR/USD exchange rate. 
 The unit of interest in this study are central concepts in public Twitter-based discussions on the Euro crisis conjecturing that this debate is a viable proxy for discussions on the EUR/USD exchange rate. 
 It is expected that some of these concepts have the potential to predict the EUR/USD exchange rate. 
  In order to test this assumption, the concepts typically used in discussions of the Euro crisis have to be identified. 
 Discussions on the Euro crisis  can be described as a system of narratives 
\citep{Propp_68} which in turn provides a heuristic approach for selecting concepts on Twitter that are possibly related to the EUR/USD crisis.

Is the focus on concepts of the Euro crisis overly specific in that other determinants of the EUR/USD exchange rate, e.g., economic problems in the US, are left out? 
Such reasoning overlooks that public discussions today take place in a competitive attention economy
\citep{Davenport_01,Falkinger_08}. 
This means that on the assumption that public global discourse is not massively manipulated, the Euro crisis or any other major theme that enters the global news agenda will be discussed relative to other topics.  Thus, even with clear focus on concepts of the Euro crisis this approach does not in itself rule out other sources of influence on the EUR/USD exchange rate. For instance, intensive global discussions of news related specifically to the US, e.g., government shutdown, may reduce the global attention devoted to the Euro crisis. Whether and to what degree an event $A$ actually takes away the public attention from event $B$ can be conceived of as the result of a global voting process which is expected to manifest itself on Twitter and other social media.

\section{Theoretical Background}
\label{Modeling}
The conceptual framework used for conducting a horse race between the efficient market hypothesis and the hypothesis of sufficient predictive information 
was the autoregressive integrated moving average (ARIMA) methodology
\citep{Box-Jenkins_76}. 
ARIMA models that correspond to the efficient market hypothesis on the one hand to the 
hypothesis and the hypothesis of sufficient predictive information on the other are 
the random walk model and regression with ARIMA errors or regARIMA model
(equations \ref{randomwalkmodel} \hspace{-0.0pt} and  \hspace{-0.0pt}\ref{regression_with_ARIMA_errors}\hspace{-0.0pt}). 
The decision between the competing models is made on the basis of in-sample information criteria (AIC, BIC)
and out-of-sample forecast error measures (MAE, MSFE)
using time series cross-validation
\citep{Arlot-Celisse_10,Hyndman_10}.\\

\subsection{Random Walk Model}
The first and simplest model to be examined is the univariate {\it random walk model} according to which a time 
series $z_t$ (or in backshift notation $B^{0}z_t)$ hinges only on its predecessor $Bz_t$ and a random process 
$a_t$. A number of authors found little evidence that  exchange rate movements (EUR/USD, ECU/USD) do not follow the random walk model 
\citep{Chen_11, Newbold+_98}.
In this study, the random walk model without drift has been chosen as it is the toughest benchmark to beat
\citep{Rossi_13, Meese-Rogoff_83}.

\begin{equation}    
	B^{0}z_t = Bz_{t} +a_{t}       
	\label{randomwalkmodel}
\end{equation}

The random walk model is a special case of the {\it autoregressive model}  abbreviated AR($p$). 
This model uses $p$-times lagged  versions of the forecast variable for prediction.
If $p$=1 and $\phi$=1, then the autoregressive model is equivalent to the random walk model

\begin{equation}    B^{0}z_t=\phi_{1} B{}z_{t} + \phi_{2} B^2{}z_{t} + \phi_{p}B^p{}z_{t}  + a_t.      \label{autoregressive_model}\end{equation}

\subsection{regARIMA Model}
The second type of model harnessed to predict and to explain the EUR/USD exchange rate is the multivariate  {\it regARIMA model}
\citep{Hyndman-Athanasopoulos_12}. 
This type of time series method incorporates one or several exogenous series  (regressors) into an ARIMA framework to assess whether the marginal explanatory power of the regressor(s) used is larger than that of a pure autoregressive model.\footnote{In this study, all $k$ regARIMA models considered after completion of the feature selection process are bivariate or uni-covariate. Each of them ties-in one exogenous time series, which is the Tweet count of one concept used in the discourse of the Euro crisis.} 
A regARIMA model 
can be conceived of as a generalization of either a regression or an ARIMA model and constructed accordingly. 
Suppose, model development starts with the regression part of a regARIMA model. Let $y_t$ be a time series of length $t$ modeled via an OLS  regression with $m$ predictor variables 
and denoted as 

\begin{equation} y_t = \sum\limits_{i=1}^r  \beta_i	x_{it} + z_{i}.   \end{equation}

The term $z_i$ refers to the residuals or regression errors, i.e., the difference between a score $y_i$ to be predicted and its model prediction with
 \begin{equation} z_t = y_{t} - \sum\limits_{i=1}^r  \beta_i	x_{it}.  \end{equation}

The regression errors $z_t$ form themselves a univariate time series which typically includes  correlated residuals. This is at odds with OLS  regression as it leads to biased standard errors and thus distorted estimates of parameters.  
At this point, the ARIMA part of a regARIMA model enters the game because correlated scores that cause problems in an  OLS regression model can well be accounted for in an ARIMA($p$,$d$,$q$) model. 
To specify the parameters of the ARIMA($p$,$d$,$q$) model such that it fits the time series of residuals $z_t$ best, there are a number of methods available like, e.g., analysis of the patterns of (partial) autocorrelation of $z_t$ or the automated procedure for optimal ARIMA selection 
\citep{Hyndman-Khandakar_08}. 
If, for instance, an appropriate model identification procedure indicates that a non-seasonal ARIMA(1,2,1) model accounts for the time series of residuals best,  then  this model model can be written in succinct backshift notation as 

 \begin{equation} (1 - \phi_{1}B) (1-B)^2 z_t = (1 + \theta_{1}B)e_t .   \label{eq4} \end{equation}

While the error term of the regression analysis $z_t$ occupies the slot of the time series, 
the error term of this ARIMA model $e_t$ is expected to be a Gaussian white noise process
\citep{Hyndman-Athanasopoulos_12}. To make visible the forecast variable $y_t$ to be modeled, 
the regressors $x_{it}$ and their (fixed) coefficient $b_i$,  the residual term $z_t$ in equation \ref{eq4} is to be replaced by  $y_{i} - \sum\nolimits_{i=1}^r  \beta_i	x_{it}$. 
Then, the regARIMA model can be re-expressed as

\begin{equation} (1 - \phi_{1}B) (1-B)^2 (y_{i} - \sum\limits_{i=1}^r  \beta_i	x_{it}) = (1 + \theta_{1}B)e_t .   
\label{regression_with_ARIMA_errors} \end{equation}

\subsection{Diagnostic Checking}
\label{DiagnosticChecking}

By its very nature, regARIMA modeling involves diagnostic tests that are motivated both by the regression part and the time series part of this model. 
Firstly, to check fulfillment of the requirements of the regression part of the regARIMA model variance was examined.
This was achieved by 
calculating variance inflation factors (VIF) 
thereby testing for collinearity of the covariates 
\citep{Fox_97}.
Low VIF scores indicate collinearity due to redundant explanatory variables, and variables with low VIF score were discarded. 
Secondly, diagnostic checking for time series was addressed. Typically, this stage of the analysis  involves a visual examination of the residuals from the tentatively entertained model as 
the distribution of residuals can reveal problems of the model applied. In this study, however, for each of the component times series  considered numerous  
time series segments (estimation windows) were analyzed. This followed from the fixed-size rolling window approach used for cross-validation (see below). Thus, visually examining correlation or partial autocorrelation plots was not an option. Instead, for each concept and for each estimation window $R$ 
the {\tt auto.arima()} function of the R forecast package
\citep{Hyndman-Khandakar_08} 
was applied as part of each of the forecasting runs conducted. This function uses unit-root tests, seasonal root tests and 
a collection of standard 
model selection criteria, e.g., AIC,  to identify the  regARIMA model parameters {\it p}, {\it d} and {\it q}  that explain the data best.

For each of the $k$ component time series analyzed the regARIMA model parameters suggested by 
{\tt auto.arima()} for each estimation window were used for model fitting. 
Likewise according to the  output of {\tt auto.arima()} each estimation segment  has been differenced, detrended and seasonally adjusted. 
To double check whether or not the adjustments made actually removed  
time series anomalies standard residual-base diagnostic tests have been carried out for each estimation window. 
These include 
the White test for nonlinearity 
\citep{Lee+_93}, 
the Ljung-Box portmanteau test to check the independence of the residuals, 
the Augmented Dickey-Fuller and the Phillips-Perron test  to examine the stationarity of the residuals via unit-root testing. 
Anomalies usually mean that the model performance is seriously affected or the model does not converge. 
In the majority of cases, the test results confirmed that the adjustments made removed the anomalies tested. 
Computation of the model prediction either via random walk model or regARIMA model 
per segment was skipped whenever an anomaly was detected. 
In this case, the calculation of mean forecast errors (MAE, MSFE) and mean information criteria scores (AIC, BIC) for a component time series $j$ was conducted without the anomalous segment. 
This applied to less than 2\% of all estimation segments examined. 
In all other cases the fitted model was  used for one-step ahead forecasting. 
For the next estimation window, 
{\tt auto.arima()}  was again harnessed to identify the ARIMA model 
unitl the end of the component time series was reached.

\subsection{Randomized Data}
What could have been achieved by just guessing randomly? Computer-generated random ``guesses'' provide reference scores 
that help to answer this question. They are essential in work on modeling as they allow modelers and the scientific community to put the results of the model tested (alternative model) into perspective. Reference scores can be derived from some sort of random model (null model), or they are based on randomized data. The random walk model introduced above is an example of the first strategy. In addition, reference scores have been generated by randomizing data. In this work, both strategies for generating reference scores have been deployed as this increases the chances of identifying false positives. 

Usage of randomized data as a validation strategy comes in many forms. Examples in point are studies that create randomized data to facilitate direct comparison with original data 
\citep[e.g., ][]{Sato-Takayasu_13}. Other approaches proceed by shuffling the original data $n$ times thereby approximating the null distributions to develop significance tests in the sense of 
\cite{Theiler+_92}. 
The work in this paper follows the first type of approach in that a random sample of the original data has been used to establish
 a  noise covariate {\it Rand} as a reference score that is instrumental in rejecting false positives
 \citep{Flack-Chang_87}. 
  The  covariate  {\it Rand}  has been generated by sampling with replacement the values (Tweet counts) 
 of all remaining 17 ``true'' covariates considered (e.g., {\it bank}, {\it risk}). As a consequence, the final calculation of error metrics (MAE, MSFE) and metrics of information criteria (AIC, BIC) 
 have been calculated for $k$ = 17 + 1 predictors or covariates.

\section{Data Selection and Feature Extraction}
\label{Data_Selection_and_Feature_Extraction}
One of the major challenges when analyzing data from the Internet, e.g., from Twitter,  
consists of selecting appropriate data and then extracting features\footnote{In this paper, the terms {\it concepts},  {\it features} and   {\it covariates} have related, though distinguishable meanings. {\it Concept} is a term used to refer to $n$-grams monitored on Twitter, e.g., ``loose spending''. More precisely, the number of Tweets mentioning one or more $n$-grams is recorded and analyzed.  The term {\it feature} is used in situations when  {\it concepts} are involved in feature extraction. {\it Covariate} is the term taken to refer to concepts as seen from the viewpoint of a regARIMA analysis.} 
correlated with or predictive of the forecast variable, e.g., the EUR/USD exchange rate. In the context of the efficient market hypothesis this challenge has been referred to as the ``messy problem of deciding what are useful information''
\citep[][p.\ 1575]{Fama_91}. 
In this study, data selection and feature extraction followed a grow-and-shrink approach of 5 steps (see Table \ref{Tab_Grow_and_Shrink}\hspace{-1.3pt}).  
Steps 1 -- 2 make use of heuristic approaches  to generate a list of \NumberofConcepts{} concepts that are plausible candidates, i.e., a long list, for 
predictors of the forecast  variable (``grow''). 
Step 3 refers to data collection, i.e., the collection of frequencies of  Tweets that mention concepts identified in the preceding step. 
Step 4 involves data preprocessing, feature selection and extraction which leads to a short list of 17 concepts. 
Step 5 identifies concepts which when used  as covariates in regARIMA models outperform the prediction of the random walk model. 
Next is a more detailed description of each of these steps. 

\begin{table}
\centering
\begin{tabular}{ccr}

	\hline
	Step & Concept Elicitation and Feature Extraction & \# \\\hline
	1 & Generating a Seed List of Concepts 										& 65 \\
	2 & Extension of the Seed List by Human Analysts 								&  \NumberofConcepts{} \\
	3  & Collection of Tweet Counts of concepts related to the Euro Crisis					 			& \\
	4 & Data Preprocessing													&  \NumberofFinalConcepts{}\\
	5 & Identifying concepts with repeatedly low forecast errors 						& 2\\
	\hline
	
\end{tabular}
\caption{Steps in concept elicitation and feature selection and number of concepts selected per step}
\label{Tab_Grow_and_Shrink}
\end{table}

\subsection*{Step 1: Generating a Seed Set of Concepts}
\addcontentsline{toc}{subsection}{Step 1: Generating a Seed Set of Concepts} 
The EUR/USD exchange rate is influenced by a large number of determinants most of which are unstable or can be identified only post hoc.  
Which of these is associated with the EUR/USD exchange rate or even predicts it?
Clearly, there is a vast number of public debates that could possibly influence the EUR/USD exchange rate. 
A plausible candidate of a public discussion that could have this potential is the global discussion on the Euro crisis. 
At the time of writing this article, the Euro crisis can be described as a narrative or a system of different and competing narratives\footnote{The British, French, German or Greek, etc.\ view of the Euro crisis.} 
with far-reaching effects on foreign exchange markets  including the EUR/USD exchange rate 
\citep{Stracca_13}. 
In this work, collecting Tweets related to the Euro crisis is based on the following assumptions:  
The intensity of the public debates on the Euro crisis is associated with or even predictive of the EUR/USD exchange rate.  
Using a keyword-based approach, narratives about the Euro crisis can be described by recurring concepts which are used on Twitter and on other media to spread information about the Euro crisis. 
 The language of the Euro crisis narratives includes concepts  like, e.g., {\it debts}, {\it protests}, {\it Germany}, {\it Greece}.  
Clearly,  any concept of a given language may become part of a Euro crisis narrative, and the language of the Euro crisis is bound to change to some degree.  
The Euro crisis can spread or change its symptoms, new financial or political threats may loom, sooner or later different politicians enter the political arena while others drop out of the public limelight, new financial instruments may be introduced or new linguistic ways of describing phenomena may emerge. But there are also concepts of the Euro crisis that seem to have a longer life-time, e.g., names of countries. It is assumed that at a given point in time,  a set of concepts of the Euro crisis with a high typicality can be defined.  
Concepts of the Euro crisis narrative(s) need to be elicited and  assembled to a seed list of concepts hypothesized to correlated with or predictive of the scalar forecast variable EUR/USD exchange rate.\footnote{In this study, only English concepts have been considered.} 

Led by the aforementioned assumptions, a heuristic selection process has been used to assemble an initial seed list of candidate concepts. The discussion of these concepts on Twitter is expected to be  predictive of the EUR/USD exchange rate. The candidate list of concepts should be large because all subsequent selection steps will lead to a reduction of this list. Errors of commission, i.e., including wrong concepts, can be fixed in subsequent steps of features selection, but errors of omission cannot. 
To compile a seed list of concepts with a predictive potential for the EUR/USD exchange rate (step 1 of Table \ \hspace{-0.03cm}\ref{Tab_Grow_and_Shrink}\hspace{-0.0009cm}) 
theories from narrative science were considered. 
The work of Vladimir Propp (1895 -- 1970) was found to be useful in achieving this purpose. 
Propp analyzed the structure of more than 100 Russian folktales and identified a set of 31 recurrent  components  which Propp called functions or narratemes 
performed by 7 dramatis personae 
\citep{Propp_68}. 
According to Propp, the characters can be either fused or spread across different persons. There can be fewer functions or narratemes but 
their sequence is always kept. 
Some of the narratemes always come in pairs, e.g., the interdiction and its violation. In addition to the narratemes, Propp suggests that there are 8 recurrent characters like the villain, the dispatcher, the magical helper and others. These structural elements have repeatedly been used to analyze  narratives in many fields as diverse as fairy tales, religious texts, political discourse
\citep{Pierce_08} and
in computer linguistics
\citep{Bod+_12}.  

Table 
\ref{Tab:Propp1} shows in prose how Propp's narratemes can be mapped to the Euro crisis. 
This reconstruction indicates that many of the historical events of the Euro crisis seem to fit Propp's narratemes. The reconstruction spans the narrative of the Euro crisis from 
{\sc absentation} (To join the common currency many European states give up their national currencies) over {\sc interdiction}  (The treaty of Maastricht stipulates that only member states with budget deficits of up to 3\% of the GDP are entitled to join the Euro) and {\sc violation of interdiction} (Many member states of the EU have budget deficits higher than
the agreed-upon 3\%) to {\sc guidance}  (Guidance and recipes how to overcome the crisis are offered by various parties).
This reconstruction did not deploy all 31 narratemes of Propp's methodology. Firstly, at the time of conducting this study the Euro crisis is still ongoing. Secondly, in a narrative analysis according to Propp not 
all narratemes have to the used. 
The narrative of the Euro crisis may be told differently in  different countries. For instance, the role of the villain in the Euro crisis narrative may be filled differently in different countries. The latter aspect is very important as a significant part of the discussion on the Euro crisis is concerned with a competition of different versions of this narrative. Still, each of these competing narrative instances complies with a Propprian analysis. One way to retell the Euro crisis in Propprian terms is to conceive of the common people in Europe as the hero. 
Table \ref{Tab:Propp2} is simply a stripped down version of Table \ref{Tab:Propp1}. Now, the Euro crisis in prose is reduced to key-words. 
The key-word based version is not committed to a particular view of any of the Euro crisis narratives.  
Whenever different instantiations of one role (e.g., hero, villain) were possible each of them was used. 
The key-word-only reconstruction of the Euro crisis narrative(s) provided  a seed list made up of 65 concepts.

\subsection*{Step 2: Extension of the Seed List by Human Analysts}
\addcontentsline{toc}{subsection}{Step 2: Extension of the Seed List by Human Analysts} 
The seed list of 64 concepts generated in step 1 covered essential concepts related to the Euro crisis. 
Intuitively, however, it seemed to be obvious that a large number of concepts was missing. 
In December 2011, a concept elicitation study was conducted to extend the seed list.  
Three male and two female native English-speaking students of computer science of the National College of Ireland in Dublin aged between 19--23 years 
took part in this study as part of their course requirements. The participants of this study were presented with the seed list of concepts generated in step 1. 
They were told that the seed list was intended to become a comprehensive concept inventory on the Euro crisis. 
But since the list was obviously incomplete the task was to provide additional concepts which could be unigrams or $n$-grams. It was emphasized that a broad spectrum of concepts was required that should tap into economic, political, social, emotional or other aspects of the Euro crisis.  The participants were asked to work individually. They were 
 free to make use of any Internet resources of their choice, and they had 90 minutes time to complete the task. The lists of concepts obtained from each student were pooled, typos were corrected and duplicates were removed. Different linguistic forms of the same concept, e.g, {\it bank\/} and  {\it banking\/} were kept, however. The abbreviation {\it SP} for 
 {\it Standard \& Poor's} was used by the subjects and hence this short form and not the long version was deployed. 
   The resulting list covered 529 concepts, which together with the 65 concepts obtained in step 1 summed up to a long list of \NumberofConcepts{} 
   concepts.  These concepts were used to elicit Tweet counts on each of these concepts per day throughout the study (Table  \hspace{-2.0pt}   \ref{Tab:Total_List_of_Concepts} \hspace{-2.0pt}).

\subsection*{Step 3: Collection of Tweet Counts of Concepts related to the Euro Crisis}
\addcontentsline{toc}{subsection}{Step 3: Collection of Tweets on the Euro Crisis} 
Concept usage on social media has repeatedly been shown to be indicative of variables of interest  both with regard to  individuals 
\citep{Choudhury+_13} 
and groups or communities   
\citep{Bryden+_13}. 
In this study, concept usage on Twitter has been harnessed as a social marker that reflects cognition and emotion related to the Euro crisis.  
From \BeginofDataRecording{} -- \EndofDataRecording{}
Tweet counts of concepts used to discuss the Euro crisis were elicited on a daily basis via R 
\citep{R_13}
and an API from the Twitter search  engine {\tt http://www.topsy.com}.\footnote{{\tt http://www.topsy.com} has  access to the full Twitter firehose.}
In this way,  \NumberofConcepts{} automated search requests have been submitted every day. 
Likewise and for the same time interval the EUR/USD exchange has been sourced in an automated way from {\tt http://www.quandl.com/.}

It is obvious that simply determining the number of Tweets 
that make use of a particular concept will generate misleading results.  For instance, if the EUR/USD exchange rate is to be examined then 
determining the number of Tweets that mention the concept {\it Euro} would return all Tweets that provide some price information in 
Euro. This type of retrieval would hardly separate Tweets that discuss the Euro currency   
 from Tweets with everyday chitchat about prices. 
This is the reason why the  \NumberofConcepts{}  concepts  assembled in steps 1 and 2 were used as selectors to target the search for Tweets on the Euro crisis. 
Hence,  each of the \NumberofConcepts{} daily requests to the Twitter search engine {Topsy} followed the pattern

\begin{center}
	Euro {\sc and}  Crisis {\sc and} $<$concept$>$.
\end{center}

 The data returned reflects the count of Tweets or Retweets that mention one or more of the 
  concepts in connection with the Euro crisis. The data does not include information on the regional origin of Tweets or the network structure among Tweeterers.\footnote{Clearly, this procedure cannot completely rule out false positives or false negatives. It is expected, however, that the large number of commercial offerings on Twitter that express prices in US Dollar or Euro should be filtered out.} 
All data elicited and used in this study, i.e., the EUR/USD exchange rate and the frequency of Tweets on the Euro crisis are time series. 
The times-series are even-spaced reflecting the rhythm of daily data collection.

Using a more formal lingo, the univariate time series on the EUR/USD exchange rate together with the $k=$ \NumberofConcepts{} 
univariate time series on the frequency of Tweets that mention one  or more of the $k$ concepts  on Twitter form a multivariate time series 
 ${\bf Y}_{1+k,t}$  from $t = \{1,2 \ldots, T\}$ 
with $j = \{0,\ldots, k\}$ univariate component time series. 
With a forecast horizon of $h=1$, the EUR/USD exchange rate was the scalar target variable  denoted by $y_{0,t+h}$ or simply $y_{t+h}$. Its forecast at time $t$ is denoted by
$f_t$ \citep{Giacomini-Rossi_13b}.
The $k$ univariate component time series $y_{j,t}$ were  the explanatory variables. 
The predictive power of each of them was assessed
via a forecasting scheme with a rolling estimation window of fixed size $R$.  Following this forecasting scheme, the total length of the multivariate time series to be analyzed $T$ was  successively split into 
in-sample portions of fixed length $R$ 
and $P$ out-of-sample portions of length $h$. For each component time series  $y_{j,t}$ the number of possible predictions $P$ can then be calculated as follows

\begin{equation}    
P \equiv T - R + 1.		
\label{number_of_predictions}
\end{equation}

\subsection*{Step 4: Data Preprocessing}
\addcontentsline{toc}{subsection}{Step 4: Data Preprocessing} 
It is obvious that not all concepts elicited in steps 1--3  have the potential to predict the EUR/USD exchange rate. To identify those concepts negative selection (rejecting inappropriate concepts) was combined with a positive selection (accepting appropriate concepts). This step was concerned with the negative selection. A number of  those concepts which initially appeared to be plausible candidates for predictors of the forecast variable turned out to be rarely used on Twitter so that their Tweet counts were low or even zero. 
These concepts can be easily detected as their variance is zero or near-zero. 
Since concepts with low variance cannot be good correlates or predictors of the EUR/USD exchange rate they were discarded.\footnote{Removal of zero- and near zero-variance predictors was achieved via the {\tt nearZeroVar()} function of the {\tt R caret} package \citep{Kuhn_13}.} 
As the Breusch-Pagan test revealed the presence of heteroscedasticity in the time series of the  EUR/USD exchange rate it was log-transformed prior to the analysis.\\

\subsection*{Step 5: Identifying Concepts with repeatedly low Forecast Errors}
\addcontentsline{toc}{subsection}{Step 5: Identifying Concepts with low Forecast Errors} 

All concepts assembled in steps 1--2 
were assessed with respect to their power to forecast the EUR/USD exchange rate.  
Good predictors ought to have a low forecast error. How low should the forecast error be to call a predictor good? To answer this question, 
 forecast errors of  the regARIMA model under study for each of  the \NumberofConcepts{} concepts have been compared with the forecast errors of a baseline model.  
In time series analysis, this baseline model is usually the random walk model. 
The forecasts have been deployed to set up a horse race between the 
hypothesis of sufficient predictive information and the 
efficient market hypothessi. While the former maintains that regARIMA models predict the EUR/USD exchange rate best, the latter claims the same of the 
random walk model. The identification of concepts with low forecast errors will be outlined in more detail in the next section.

\section{Results}
\label{Results}
\subsection{Data}
From \BeginofDataRecording{} -- \EndofDataRecording{}, i.e., for \DaysofDataRecording{} days{\footnote{At this time,  {\tt topsy.com} changed the terms of trade and introduced some technical changes to the API.}}, Tweet counts on the Euro Crisis were garnered from 
{\tt topsy.com}, a reseller of Twitter data. The EUR/USD exchange rate was recorded for the same time span. 
The Tweet counts reflect the number of Tweets mentioning the \NumberofFinalConcepts{} selected concepts on the Euro crisis. 
Accordingly, this step generated \NumberofFinalConcepts{} time series. 
After preprocessing these  time series the initial long list of 
\NumberofConcepts{}  
concepts was reduced to the following short list of \NumberofFinalConcepts{} concepts and their associated time series. 

\begin{quote}
  bank,   banking,  banks,  debt,  ECB,  economy,  Euro,   Germany,  Greece,  Greek, Hollande, Italian, Italy, Moodys, risk, SP, Spain   
\end{quote}

\begin{figure}[h!]
 \centering
     \includegraphics[width=1.0\textwidth]{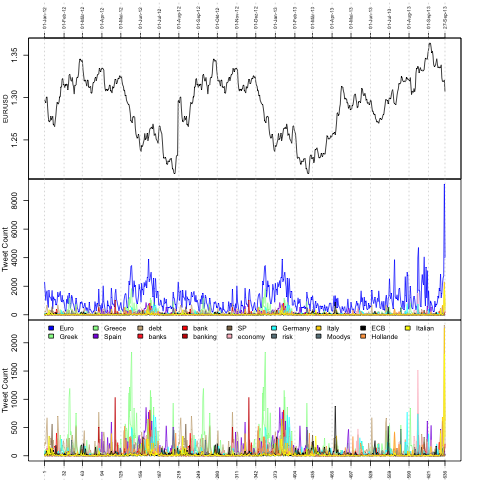}
    \vspace{-8.8pt}
      \caption{EUR/USD exchange rate  Jan 1st 2012 -- Sep 27th 2013} (top) and Tweet counts including the  concept {\it Euro}  (middle) and without it (bottom) 
            \label{Dollar_Euro_Tweet_Counts}
\end{figure}

A descriptive account of the time series of Tweet counts along with the EUR/USD exchange rate is given by Figure \ref{Dollar_Euro_Tweet_Counts}. 
The three panels of this figure use a common time axis to align the  EUR/USD exchange rate and the number of Tweets that made use of one or more of the shortlisted 17 concepts. 
The top panel of this figure presents the EUR/USD exchange rate. The remaining two panels show the Tweet counts collected during the same time span. 
All panels of this figure use a common time axis to align the  EUR/USD exchange rate and the number of Tweets that mentioned at least one of the shortlisted 17 concepts. 
The middle panel presents Tweet counts of all 17 concepts, the lower panel features as well Tweet counts but leaves out the Tweet counts for Euro. 
 Not surprisingly, the concept {\it Euro} was used most often (blue series in the middle panel of Figure \ref{Dollar_Euro_Tweet_Counts}). As this series eclipses other series it has been left out in the lower panel. Prima facie, both panels on Tweet counts suggest  that a downward  (``bearish'') trend of the EUR/USD is associated with an increased discussion of the Euro crisis as evidenced by peaks in the usage of the concepts {\it Euro} and {\it Greece}  around the middle of June 2012 and middle of December 2012.  This type of eyeball econometrics is  limited. For instance, just focusing on some salient feature means that the overall data base is bound to be thin. The following sections will use a different approach to data analysis and prediction.

 \def\ignore#1\endignore{}

\newcolumntype{H}{@{}>{\ignore}l<{\endignore}}

\subsection{In-Sample Model Selection}
The research question of this paper asks whether  and under what conditions the hypothesis of sufficient information
that operates on the basis of data gleaned from Twitter or  the efficient market hypothesis facilitates better predictions of the EUR/USD exchange rate. 
This question can be rephrased as a selection between these two classes of models and also as a selection among uni-covariate regARIMA models that outforecast the random walk model.
n work on time series model selection, information theoretic approaches play a dominating role, and it is this class of methods that will be considered here.  
Methodological debates in time series model selection often revolve around the question whether Akaike's information criterion 
\citep[AIC,][]{Akaike_74}, the Bayesian information criterion 
\citep[BIC,][]{Schwarz_78} 
or other information criteria are most appropriate for a given model selection problem.
It is well documented that AIC has the tendency to over-fit and to under-penalize complex models while  BIC makes use of a stricter penalty regime for additional parameters. Hence, in comparison to AIC it is BIC that tends  to prefer simpler and more parsimonious models.   
In this study, the models examined are nested which involves, necessarily, issues of model complexity. 
Comparing and contrasting both AIC and BIC scores is useful when deciding whether the additional complexity of the regARIMA models leads to different results of the model comparison or whether 
despite this, AIC and BIC scores agree. 

Against the backdrop of the discussion above, both in-sample AIC and BIC scores have been calculated  for all models considered.  
The scores provided the basis for a ranking of all models or covariates.
Table    \ref{InformationCriteriabyPredictorsAveragedacrossEstimationWindows} 
presents the scores of the averaged scaled information criteria used for the random walk model and the regARIMA models.
All scores have been rescaled which is denoted by $\Delta$AIC and $\Delta$BIC.  This means scores of BIC and AIC are expressed relative to their 
minimum value found in the overall set of random walk model and the regARIMA for all $k$ predictors and estimation windows.

\begin{table}[ht]
\begin{small}
\centering
\begin{tabular}{lcccc} \hline
 &  \multicolumn{2}{c}{$\Delta$AIC} & \multicolumn{2}{c}{$\Delta$BIC} \\\cline{2-5}        		
Predictors  & random walk &regARIMA&random walk&regARIMA\\ \hline
bank & 352.13 		& 4.55 & 360.83 	& 4.98 (12)\\ 
banking & 352.13 	& 5.25 & 360.83 	& 5.54 (15)\\ 
banks & 352.13 	& 4.14 & 360.83 	& 4.42 (08) \\ 
debt & 352.13 		& 0.00 & 360.83 	& 0.00 (01) \\ 
ECB & 352.13 		& 2.18 & 360.83	 	& 2.66 (04) \\ 
economy & 352.13 	& 4.60 & 360.83 & 4.89 (17)\\ 
Euro & 352.13 		& 2.82 & 360.83 	& 3.27 (05) \\ 
Germany & 352.13 	& 4.26 & 360.83 & 4.56 (09)\\ 
Greece & 352.13 	& 5.22 & 360.83 	& 5.51 (14)\\ 
Greek & 352.13 	& 5.32 & 360.83 	& 5.61 (16)\\ 
Hollande & 352.13 	& 5.16 & 360.83 & 5.45 (13)\\ 
Italian & 352.13 	& 3.99 & 360.83 	& 4.19 (07)\\ 
Italy & 352.13 		& 4.28 & 360.83 		& 4.58 (10)\\ 
Moodys & 352.13 	& 1.61 & 360.83 	& 1.86 (03)  \\ 
rand & 352.13 	& 1.23 & 360.83 	& 1.52 (02) \\ 
risk & 352.13 & 3.83 & 360.83 		& 5.84 (18)\\ 
SP & 352.13 & 3.09 & 360.83 		& 3.40 (06) \\ 
Spain & 352.13 & 4.50 & 360.83 	& 4.79 (11)\\ \hline

\end{tabular}
\caption{Information criteria by predictors averaged across estimation windows ($R$ = 310 -- 620)} 
\label{InformationCriteriabyPredictorsAveragedacrossEstimationWindows}
\end{small}
\end{table}

{\it Differences between random walk model and regARIMA models. } The random walk model was clearly outperformed by all regARIMA models. 
In other words, when integrated into uni-covariated regARIMA analyses, each concept (predictor, covariate) listed in Table  \ref{InformationCriteriabyPredictorsAveragedacrossEstimationWindows}
or rather its associated Tweet count time series facilitated better $\Delta$AIC and $\Delta$BIC scores than the random work model. 
It is unlikely that this result is attributable to the higher model complexity of regARIMA models as it was consistently found for AIC and for BIC.

{\it Comparison of AIC and BIC. } In line with the discussion on information criteria, AIC scores were lower than BIC scores indicating a better model fit which, however, may be due to the smaller penalty that AIC uses to castigate model complexity. 
For the random walk model, AIC and BIC provided uniform values as covariates are by definition ignored in this class of model. 
With respect to the regARIMA models, separate rankings for AIC and for BIC scores could be established the positions of which indicated in brackets in Table  \ref{InformationCriteriabyPredictorsAveragedacrossEstimationWindows}. 
While AIC and BIC agree on the 5 best explaining concepts (debt, Moody's, ECB, Euro, SP), the ranking positions of AIC and BIC disagreed 
(or agreed again) with regard to the remaining ranking positions.

{\it Comparison among Covariates.} The covariate {\it debt} showed a better performance than all other covariates. This performance gain was consistently expressed in terms of $\Delta$AIC 
and $\Delta$BIC and at the same time this performance gain was consistent across all estimation windows considered. The noise covariate {\it Rand} came second which means that all predictors 
but {\it debt} performed below chance level. This is a remarkable finding. It indicates that using error scores of the random walk model as baseline values appears to be a too liberal approach which may contaminate the results with false positives.

\subsection{Out-of-Sample Forecasting}
Forecasting and evaluating the accuracy of the forecasts are essential parts of predictive modeling. The accuracy is higher the smaller the difference between the realization of the forecast variable $y_{t+1}$ 
at time $t$ and its forecast $f_t$. This intuition is typically expressed by a loss 
function $L(\cdot)$. In the majority of studies that involve time series forecasting 
the quadratic loss function (mean square forecasting error, MSFE) or  the absolute loss function (mean absolute error, MAE) is harnessed. 
Accordingly, the accuracy of a forecast can be described as the expected loss $E[L(y_{t+1},f_t)]$. The latter is usually operationalized by 
averaged pseudo out-of-sample forecasting error scores using MAE, MSFE  or other forecast error metrics  
\citep[e.g., ][]{Giacomini-Rossi_13b}. Both MSFE and MAE are then used to express the forecast accuracy of models of interest, e.g.,  regARIMA or random walk model. 
In time series analysis it is well known that large estimation windows are often necessary to average out measurement errors. Clearly, this is only true if  there are no structural breaks the risk of which increases with the length of estimation windows considered. In this study, medium to large estimation windows and thus small sample split values have been used.
This choice was  motivated by the assumption that time series of Tweet counts can be expected to be a noisy type of data.
Starting with $R = 310$ which corresponds to a sample split of $P/R \approx 1$ 
different forecasting runs were conducted each time increasing the size of the estimation window by ten.
With \DaysofDataRecording{} days of data recording 								% 636 
this led to a set-up with $R \in  \{310, 320, \ldots, 620\}$
resulting in s=32 sample splits and thus 32 forecasting runs with $1 \gtrsim P/R \gtrsim 0.1 $. 
Clearly, this is an explorative approach. It can be justified, though, by the novelty of time series analysis on Tweet counts. Moreover, results of all 32 estimation window sizes examined are reported here so that the robustness of the findings can be assessed. 

\begin{figure}[h!]
 \centering
 \hspace{150.0pt} 
     \hbox{\hspace{9ex}\includegraphics[width=1.0\textwidth]{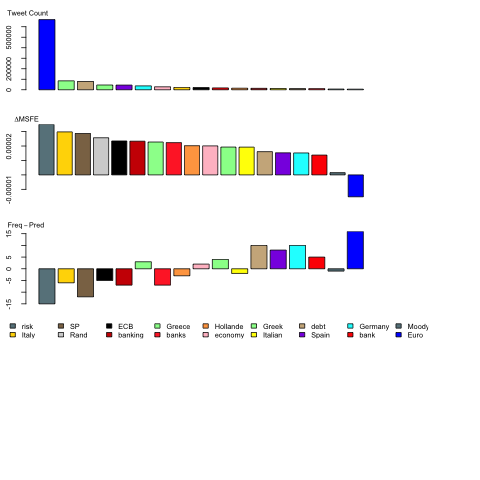}}
    \vspace{-121.0pt}
      \caption{Frequency (top) and predictability (middle) and ranking differences of frequency and predictability of Tweets per concept (bottom)
            \label{Most_Frequent}}
\end{figure}

Each of the following two sections report on a different rationale to examine the predictability of covariates. 
Section \ref{PredictabilityofeachCovariatebyaveragingacrossSampleSplits} reports on a scoring approach to predictability, while
section \ref{PredictabilityofeachCovariatebycountingbetterthanRandomWalkPerformance} informs about a classificatory approach to examine the predictive power of the 17+1 covariates under scrutiny.

\subsubsection{Predictability of Covariates by Averaging across Sample Splits}
\label{PredictabilityofeachCovariatebyaveragingacrossSampleSplits}
What is the average predictive strength or predictability of each the 17+1 covariates?  
How do the predictability scores of the 17+1 covariates compare to each other, and what is the relationship between the frequency and the predictability of the 17+1 covariates considered? 
The analysis described in this section makes use of a scoring approach to answer these questions.

The results of this analysis  are shown in Figure \ref{Most_Frequent}.
It presents the frequency of Tweet counts (top), the predictability of each covariate averaged across sample splits (middle) and the frequency of covariates (concepts) relative to the 
predictability of covariates. 
The top panel of Figure  \ref{Most_Frequent}  presents the Tweet counts of the 17 concepts considered as seen from a cross-sectional angle. Thus, it provides an alternative perspective on the longitudinal arrangement of the same data in Figure \ref{Dollar_Euro_Tweet_Counts}.\ The noise variable {\it Rand} does not enter this panel as this variable is synthetic and has not been recorded. 
For each of the 17+1 covariates considered the middle panel of Figure \ref{Most_Frequent}  expresses the predictive power relative to the predictability of the random walk model averaged across sample splits.  Here, {\it Rand} has been included. To ease readability $ \Delta{}\textrm{MSFE}$  is  used to refer to the predictability scores averaged across sample splits.\footnote{More formally, let $\Delta{}$MSFE$_{jv}$ denote the predictability of a covariate $j$ of a regARIMA model relative to the corresponding random walk model for a particular sample 
split $v$. Then, the MSFE score of a covariate $j$ averaged across all $s$ sample splits under consideration can be expressed as 
\vspace{-0.5cm}

\begin{equation}
 \Delta{}\textrm{MSFE} = 
 \frac{1}{s}  \sum\limits_{i=1}^s  \Delta{}\textrm{MSFE}_{jv}.
\end{equation}}
The middle panel shows that among the 17+1 covariates analyzed the covariates {\it risk},  {\it Italy} and  {\it SP} have the highest predictability scores when used in regARIMA analyses. The finding that {\it Rand} occupies the 4th predictability rank position means that the predictive power of all remaining covariates is negligible.

The bottom panel of Figure \ref{Most_Frequent} integrates the information of the two panels above by presenting the frequency of covariates (concepts) relative to the predictability of covariates. 
Comparing frequency and predictability ranks of Tweet counts helps to answer the question whether and to what degree concepts were ``overtalked'' or ``undertalked''  on Twitter
within the context of the Euro crisis and using the methodological setup described. 
This has been facilitated by firstly ranking all covariates in terms of their frequency and then in terms of their predictability. Secondly, for each concept the 

\begin{equation}
\textrm{{\sc frequency rank position - predictability rank position}}
\end{equation}

\noindent has been calculated. For instance, the concept {\it risk} has the predictability rank position 1 and the frequency rank position 16. 
Therefore, for {\it risk} the predictability relative to its frequency is  -15.  In this sense, {\it risk} was found to be  ``undertalked'' on Twitter.   
The same can be said about all other covariates that beat both the random walk model and the noise covariate {\it Rand} (i.e., {\it risk},  {\it Italy} and  {\it SP}). Each of them is 
``undertalked'' on Twitter. 
 By contrast, for the concept or covariate {\it Euro} the frequency relative to the predictability is 17-1=16. The positive value of this score
  expresses that {\it Euro} is ``overtalked'' on Twitter. This means that the frequency of using {\it Euro} on Twitter is disproportionally high
relative to its very low predictive power to actually forecast the EUR/USD exchange rate.

\subsubsection{Predictability of Covariates that outforecast the RWM and the Noise Variable}
\label{PredictabilityofeachCovariatebycountingbetterthanRandomWalkPerformance}
\begin{figure}
\centering
\subfigure[Outperforming the Random Walk Model and {\it Rand} -- Mode Vote by Covariates]
{\includegraphics[width=12.12cm]{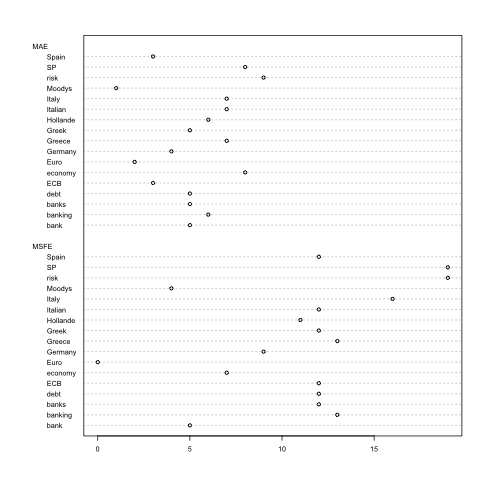}}
\subfigure[Outperforming the Random Walk Model and {\it Rand} -- Mode Vote by Estimation Window Size]
{\includegraphics[width=12.1cm]{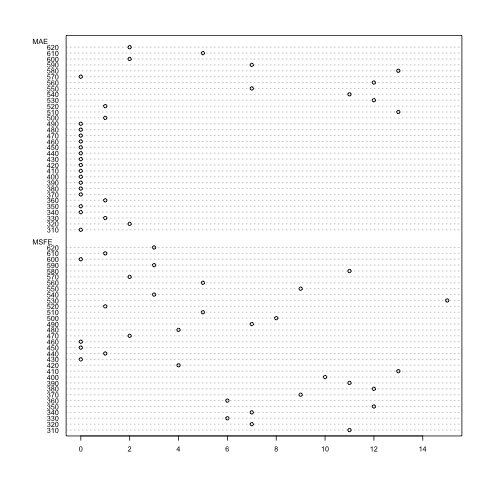}}
\caption{Mode vote of regARIMA models by covariates (a) and by estimation window size (b) }
\label{Prediction_Errors_by_Predictors_and_by_R}
\end{figure}

While the study of predictability  delineated in the previous section followed a scoring approach, the analysis described in this section pursued a classificatory approach. 
In a nutshell, the rationale of this analysis is 
(i) to classify each covariate as to whether or not it has outforecasted both the random walk model and the noise covariate
{\it Rand} and then 
(ii) to count the number of times each of the 17+1 covariates and the $s$ estimation window sizes facilitated this performance. 
This calculation provides the basis for a mode vote that indicates which covariate and which estimation window size has secured the best forecasting results. 
Figure \ref{Prediction_Errors_by_Predictors_and_by_R} illustrates the results of this analysis. 
The results are broken down for covariates  
(Figure \ref{Prediction_Errors_by_Predictors_and_by_R}a)
and estimation window sizes
(Figure \ref{Prediction_Errors_by_Predictors_and_by_R}b). 
In turn, each of the two dot plots of this figure expresses the forecast error via MAE and MSFE. 
The x-axis in Figure \ref{Prediction_Errors_by_Predictors_and_by_R}a informs how often out of 32 runs each covariate outforecasted the random walk model and the noise covariate. 
Accordingly, the x-axis in Figure \ref{Prediction_Errors_by_Predictors_and_by_R}b indicates for each of the $s$ estimation window sizes studied how many out of \NumberofFinalConcepts{} predictors outforecasted both the random walk model and the noise covariate {\it Rand}.

% ----------------------------------------------------------------------------------------------------------------------------------------------------------------------------------------------------------------------------------------------------------------------------------- %
{\it Differences between MAE and MSFE. } 
Even a casual look at the results of Figure  \ref{Prediction_Errors_by_Predictors_and_by_R} reveals differences between the results expressed by MAE 
on the one hand and MSFE on the other.  
Often, when using MSFE as a metric for forecast errors a shorter estimation window is required to identify predictors that outperform the random walk model. 
One reason for the differences is that the MSFE is more influenced by large error scores than the MAE.  
This difference becomes more expressed when the estimation window $R$ is small as this typically leads to higher error scores.

{\it Mode vote by covariates. } 
The winners of the mode vote by covariates were the predictors {\it risk} and  {\it SP} if the focus is on MSFE (Figure \ref{Prediction_Errors_by_Predictors_and_by_R}a).
Each of the regARIMA models that uses the Tweet count of {\it risk} or  {\it SP} as a covariate has outforecasted the random walk model in 19 out the 32 runs per estimation window calculated which amounts to a hit rate of 59\%. The covariate  {\it Euro} never outforecasted the random walk model and the noise covariate {\it Rand}. 

{\it Mode vote by estimation window size. } The findings reveal that when analyzing time series of Tweet counts large estimation windows were indeed required to identify predictors that outperform the random walk model. There is, however, not a simple linear relationship between the size of the estimation window and forecast accuracy. The best result in terms of MSFE was achieved with an estimation window size $R$ of 530 days. Since the total length of the time series studied was \DaysofDataRecording{} this amounts to a sample split of $P/R = 106/530 = 0.2$.

\subsection{Residual Diagnostics}
Time series diagnostics has  been conducted on the fly as an integral part of each the forecasting runs conducted for each estimation segment. 
According to the results of diagnostic tests conducted, each segment has been differenced, detrended, seasonally adjusted or in case of serious anomalies discarded (cf. section \ref{DiagnosticChecking}). 
This  section informs about additional diagnostic sanity checks. The focus is on out-of-sample final model residuals of the fitted regARIMA models  that incorporated the covariates {\it SP} and {\it risk}, respectively,  
which were examined for all $s$ sample splits or time series runs. 
The analysis of the residuals included 
the calculation of the first four standardized central moments  of the residuals to assess their distributional properties, 
Q-Q plots to facilitate assessment of normality of residuals,  
ACF and PACF plots that illustrate the correlative properties of the  residuals 
and residual tests for autocorrelation of the residuals (Box-Ljung portmanteau test),  for trend stationarity (KPSS  test) and for nonlinearity (White Neural Network Test). 

{\it Residual Moments. }
Tables \ref{Moments_SP}--\ref{Moments_risk} present  the first four standardized central moments calculated for $s$=32 estimation window sizes or sample splits. 
Both for the covariates {\it SP} and {\it risk} the values obtained for the first moment were small. This is true for the standardized values reported but applies also to the absolute values of the first moment. 
Positive and negative values of the first moment are reasonably well distributed across the sample splits studied. There is some variation of the first moment across the different sample splits which applies 
to {\it SP} and to {\it risk}. Results for the second moment exhibit only small variation, and thus there is no obvious indication for heteroscedasticity of the residuals. With decreasing sample-split ratio the values obtained for skewness turn increasingly more negative. The consistency of this trend with decreasing sample split both for {\it SP} and {\it risk} seems to suggests that regARIMA model forecasts have a tendency 
to consistently overpredict whenever the values for $R$ are large and values for $P$ are small.  However, this interpretation is  not supported by the results obtained for the first moment. Any constant trend to generate forecasts systematically higher than the actual values of the EUR/USD exchange rate would have led to consistently negative residual average scores. But this is not the case. Still, it seems that occasionally there is some emphasized overprediction. This is reflected by the negative skewness scores which is usually averaged out. Finally, both for {\it SP} and for {\it risk} the values of the fourth moment are on  moderate levels of magnitude. They exhibit  a gentle increase of the magnitude of kurtosis with decreasing sample split ratio (larger $R$) indicating a tendency towards fat tails of the residual distribution.  

\begin{figure}[h!]
     \hbox{\hspace{-13.5ex}\includegraphics[width=1.3\textwidth]{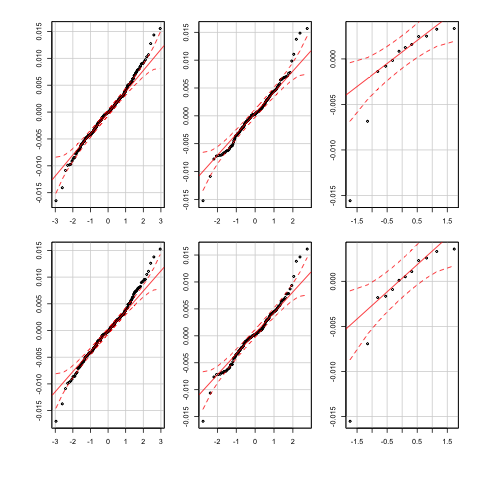}}
      \caption{Residual Q-Q plots of the covariates {\it SP} (top row) and {\it risk} (bottom row) across different sample splits with added dotted lines indicating upper and lower 95\% approximate confidence intervals. 
       Left column $R$=310, $P$=322, 
       middle column $R$=460, $P$=172,  
       right column $R$=620, $P$=12
            \label{ResidualQ-QPlots}}
\end{figure}

{\it Q-Q Plots. } Figure \ref{ResidualQ-QPlots} shows the residual Quantile-Quantile plot of the covariates {\it SP} (top row) and risk (bottom row) across three different sample splits (left, middle, right column). 
The Q-Q plots of the left column of Figure \ref{ResidualQ-QPlots} ($R$=310, $P$=322)
illustrate the situation when a comparatively small estimation window is used to forecast a large number of data points. 
Even though the data points at each end of the straight line depart to some degree  from the line, most of the data is  within the 95\% confidence bands. 
Overall, both Q-Q plots of the left column indicate that the residuals  can be reasonably accounted for by a normality distribution. 
The Q-Q plots of the middle column  of Figure \ref{ResidualQ-QPlots}  
($R$=460, $P$=172) exhibit some increased scatter around the reference line. But on the whole, the two Q-Q plots of the middle column are similar to the ones 
of the left column which have been generated on the basis of a much larger data set.  
The Q-Q plots of the right column  of Figure \ref{ResidualQ-QPlots}  
($R$=620, $P$=12) have reference lines that cut the x-axis higher than their counterparts in the four remaining Q-Q plots. This reflects the somewhat higher average score of this 
small group of residuals. Most of the residuals are on the reference line or fall within the 95\% confidence bands.
In brief, the distributional pattern of residuals does not change drastically across the 3 sample splits studied. Deviations from the straight line are small so that the Q-Q plots are 
reasonably in line with the normal distribution. This interpretation is consistent with the small scores of the standardized central third and fourth moments obtained for the corresponding sample splits.

\begin{figure}[h]
 \centering									   
     \hbox{\hspace{-13.5ex}\includegraphics[width=1.2\textwidth]{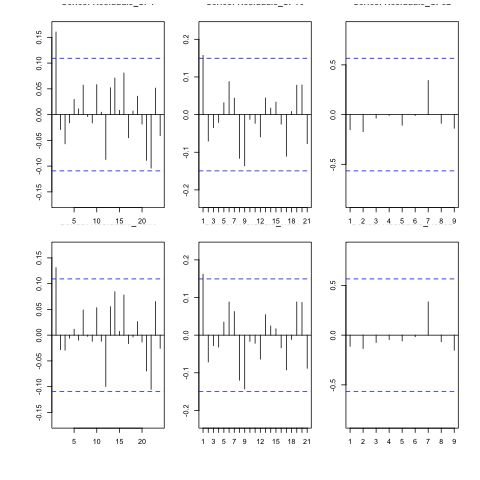}}
     \vspace{-1.0cm}
      \caption{Autocorrelation plots of covariates {\it SP} (top) and {\it risk} (bottom) across different sample splits. 
       Left column $R$=310, $P$=322, 
       middle column $R$=460, $P$=172,  
       right column $R$=620, $P$=12. Horizontal dotted lines indicate upper and lower 95\% approximate confidence intervals. 
            \label{SP_risk_for3R_Acf}}
\end{figure}

\begin{figure}[h]
 \centering
     \hbox{\hspace{-13.5ex}\includegraphics[width=1.2\textwidth]{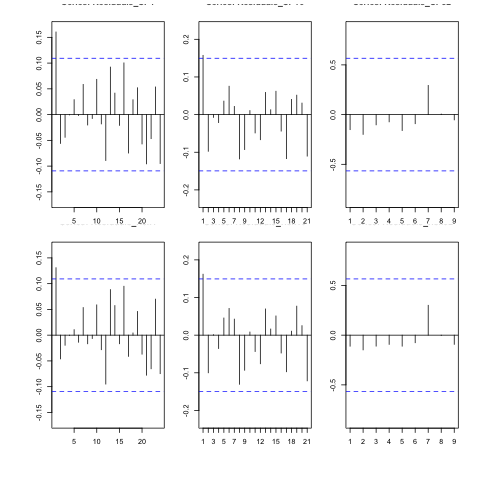}}
      \vspace{-1.0cm}
 \caption{Partial autocorrelation plots of covariates {\it SP} (top) and {\it risk}  (bottom) across different sample splits. 
       Left column $R$=310, $P$=322, 
       middle column $R$=460, $P$=172,  
       right column $R$=620, $P$=12. Horizontal dotted lines indicate upper and lower 95\% approximate confidence intervals. 
            \label{SP_risk_for3R_Pacf} }
\end{figure}

{\it Residual ACF and Residual PACF Plots}. 
At this stage of the analysis, the goal of interpreting the ACF and PACF plots was not not to identify regARIMA model parameters but to examine whether the selected residuals show signs of anomaly. 
Figure \ref{SP_risk_for3R_Acf}  presents the sample autocorrelation plots of the residuals that result from the regARIMA analysis with {\it SP} and {\it risk} as covariates. 
Figure \ref{SP_risk_for3R_Pacf} shows the corresponding partial autocorrelation plots. 
Note that the ACF plots do not include the redundant spike at lag 0. 

For both {\it SP} and {\it risk} the ACF and the PAC plots of the left column ($R$=310, $P$=322), i.e., at relatively high sample splits, reveal  a non-zero spike at lag 1.   
Apparently, the residuals  analyzed are not white noise 
which suggests that the model parameters chosen in this bracket of $P$ and $R$ values do not fit the data well. As the sample split decreases the residuals become well behaved. This is illustrated by the middle and the right columns of Figures  \ref{SP_risk_for3R_Acf} and  \ref{SP_risk_for3R_Pacf}. Here,  the non-zero spikes disappear or reach just above the confidence bands and the plots 
gives no indication  of stationarity of the residuals like, e.g., slow decay of autocorrelations.

{\it Residual Tests}. Tables \ref{Residual_Diagnostics_SP} -- \ref{Residual_Diagnostics_risk} report on the results of the univariate residual tests conducted. These tests include 
the Ljung-Box Test for autocorrelation (LB), 
the KPSS test for trend stationarity (KP) and 
the White neural network test for nonlinearity (WH). 
For each residual test, $s$ null hypotheses have been tested. 
To address the problem of multiple testing the raw $p$-values have been adjusted via the Bonferroni procedure. Both the raw and the Bonferroni adjusted $p$-values have 
been made accessible in tables  \ref{Residual_Diagnostics_SP} -- \ref{Residual_Diagnostics_risk}.

\begin{quote}

{\it Ljung-Box Test.} The Ljung-Box test examines the null of independence in the time series studied. Applied to  the analysis of residuals,  a small $p$-value indicates 
that the residuals are not independent and possibly correlated. A number of  observed $p$-values were below 5\%.  This was in particular true in a situation of relatively high sample splits. 
After applying the Bonferroni procedure the overwhelming majority of the adjusted $p$-values was higher than 5\%. Hence, for most of the sample splits the null of independence could not be rejected. 
Using alternative controlling procedure, e.g., Holm's procedure, changed this outcome only marginally.  

{\it KPSS Test.} The KPSS test has been taken to examine the null that the time series of residuals for each value of R are stationary.
A high $p$-value is a lack of evidence that the residuals are not stationary. 
The observed (uncorrected) $p$-values were never lower than 5\% so that the null of level stationarity can not be rejected.

{\it White Test.} The White neural network test for neglected nonlinearity \citep{Lee+_93} has been harnessed to test the null of linearity in mean. 
 For {\it SP} and for {\it risk} and across all sample sets the results gave no indication to reject the null. 

\end{quote}

\noindent
In short, the examinations of the residuals diagnostics characterized the conditions of good regARIMA model performance relative 
to the random walk model. While most of the residual tests conducted indicate that the requirements to ensure model adequacy are fulfilled, the ACF and PACF plots suggest that caution should be taken with short estimation windows.

\section{Conclusion}
\label{Conclusion_and_Future_Work}
``Does anything forecast exchange rates'' 
\citep{Rossi_13}? According to the efficient market hypothesis 
\citep{Fama_70, Fama_91}
the answer is a resounding ``No''.  
For some researchers this hypothesis has empirical support that is ``virtually unparalleled in economics'' \cite[][p.\ 334]{Geweke-Feige_79}. Early on, however, and even more so in recent times other scientists have critiqued the efficient market hypothesis from different vantage points 
\citep[e.g., ][]{Grossman-Stiglitz_80}. 
For the purpose of this study it may be useful to remember that the efficient market hypothesis has been developed in a more traditional ecosystem of markets and media.
 Compared to the 2nd half of the last century,  markets and media today rely on a much more sophisticated technical infrastructure. The change of the technological basis of market and media means faster information spreading and enormous speed gains in trading. It  is an open question which market mechanisms have not only speeded up but changed inherently. The concept of information is a candidate for the latter.  Even though the efficient market hypothesis is defined relative to an information set, this aspect has usually been underspecified. Moreover, its possible dependence on the technology 
  at a given historical point in time remains largely undiscussed.   
  Today`s ecosystem of markets and media adds a question mark to the simple dichotomous distinction between ``publicly available information'' on the one hand and ``insider information'' on the other which seems to be essential to the efficient market hypothesis. For instance, how do insights gained from big data analyses fit into this conceptual framework? The data that feed into such  analyses may be publicly available and the results may simulate or even go beyond the information of insiders. 

It is against this backdrop of considerations around information and markets that this study has been conducted. Information gleaned from Twitter is harnessed here as publicly available information.  Even though the representatives of efficient market hypothesis of the sixties  and seventies may not have imagined this type of information, it qualifies as publicly available information because today it can be accessed by all market participants. 
Is Twitter efficient in that it provides marginally predictive information not or not yet reflected by the EUR/USD exchange rate? Or is the market efficient in the sense that all market-relevant information are already priced-in the EUR/USD exchange rate? If the FX exchange market is a semi-strong informationally efficient market, then information secured from Twitter should not have a  marginal predictive content relative to the random walk model. The main result of this study challenges the efficient market hypothesis as publicly available information gleaned from Twitter can be shown to predict 
repeatedly the  EUR/USD exchange rate above chance level. The boundary conditions under which this result has been achieved merit particular consideration. 

{\it Selection of Tweets. } As described above, a 2-step selection of tweets proved successful in identifying covariates that predicted the EUR/USD exchange rate above chance level. 
The covariates  {\it risk} and {\it SP} have been found to out-forecast repeatedly the random walk model and the noise covariate. The semantic content of these words suggests that talking of risk on Twitter can be harnessed to forecast the EUR/USD exchange rate. Apart from the meaning of {\it risk} and {\it SP}  the result is encouraging in the sense that these concepts are not short-lived like, e.g., names of politicians many of whom play role in the public discussion for only a relatively short length of time. 

{\it Data Quality. } Data secured from Twitter is not standard in econometric time series analysis. With regard  to the covariates {\it risk} and {\it SP} the data quality of this data was on a level required for meaningful time series analysis as evidenced by most of the diagnostic tests carried out. The residual tests conducted for these covariates gave no indication that the residuals are not independent, that the data is not level stationarity or that the data is not linear. The corresponding Q-Q plots indicate that the residuals are reasonably in line with the normal distribution. 

{\it Parameter Settings. } The ratio between the number of possible predictions and the size of the estimation window (sample split) turned out to be the most influential parameter in this study. 
For predictive effects to be detected  large estimation window were required which was instrumental in bringing down the sampling error given the noisy data of Tweet counts.

{\it Safeguarding against Chance Results. } The ubiquitous modeling practice of using just the random walk model turned out to be insufficient to guard against chance findings.
To achieve this purpose it became evident that both a benchmark model, i.e., the random walk model and benchmark data, i.e., the noise covariate {\it Rand}
had to be used. With regard to the assessment of model fit, all 17 covariates outperformed the random walk model but only the covariate {\it debt} outperformed also the noise covariate. 
When it comes to out-of-sample forecasting, only the regARIMA models that outperformed  the corresponding random walk model and  the covariate {\it Rand} can be considered as evidence of a predictive effect above chance-level. This turned out to be true for {\it SP} and {\it risk} in the majority of cases (sample splits) considered.  

In summary, this study has provided supporting evidence for the hypothesis of sufficient predictive information. Under the conditions discussed above the EUR/USD exchange rate proved  forecastable above chance level using Tweet counts. The study confirms and extends the findings of \cite{Papaioannou_13}  who provide evidence that information gleaned from microblogging platforms such as 
Twitter can enhance forecasting efficiency  of intraday exchange rates. The fact that the work introduced in this paper and the study of \cite{Papaioannou_13} agree considerably in their findings 
despite using different methodologies attests to the robustness of the results.  
Even though a somewhat unlikely couple at first blush, the analysis has revealed that efficient market hypothesis and the analysis of public discussions on Twitter, illuminate each other. Seen from the viewpoint of efficient market hypothesis, the analysis of data secured from Twitter helps to identify information which may predict  the FX market.  In doing so, the methodology introduced in this article, contributes {\it not} ``to sidestep the messy problem of deciding what are useful information'' \citep[][p.\ 1575]{Fama_91}. Seen from the vantage point of the notoriously theory-poor but  data-rich area of Twitter analysis, the study has shown how efficient market hypothesis  can be harnessed to gain  insights from data on social media. 
Clearly, more research is required to further examine the relationship between public communication on social media and exchange rates. 
There are good chances that this will pave the way to apply other econometric concepts, e.g., Granger causality analysis, to  data harvested from social media, thereby contributing to a better understanding of social and economic phenomena.

\newpage
\bibliography{/Users/dietmarjanetzko/Documents/Dropbox/APA-LATEX/Library.papers3/everything}
\bibliographystyle{apa}

% ------------------------------------------------------------------- Apendix -------------------------------------------------- #

\pagebreak
\appendixpage
\addcontentsline{toc}{section}{\appendixname}

\begin{table}[hbt]
 \begin{tabular}{ccl}

 No & Narrateme                        & Euro Crisis  \\  \hline
  1 & Absentation                        & To join the common currency many European states give up their  \\
     &                                               & national currencies\\
  2 &  Interdiction                         & The treaty of Maastricht stipulates that only member states with budget          \\
     &                                               & deficits of up to  3\% of the GDP are entitled to join the Euro.   \\
  3 &  Violation of Interdiction   &  Many member states of the EU have  budget deficits higher than\\
     &                                               & the agreed-upon 3\% \\

4  &  Reconnaissance               & Financial markets monitor the economic situation of the states of the Euro  \\
    &                                                &  zone.\\

5 &  Delivery                                & Banks receive information about the sorry condition of the economy  \\
    &                                                &  in some  states of the Euro zone by ratings.\\
   
6  &  Trickery                                & Financial markets and their helpers   enforce austerity measures  \\
    &                                                & which are basically used to finance the banking system. \\

7  &  Complicity                           & Large parts of the people of Europe and their governments are deceived  \\
    &                                                & by the banks, and some are unwittingly helping them.  \\

8 &  Villainy                                & The banks and their helpers consolidate their influence. Austerity, lack of   \\
    &                                               & growth and unemployment are the consequences.\\

9 &  Mediation                          & The bad situation is made public, it is debated in the media and on EU summits. \\
    &                                                &  \\

10 & Beginning  counteraction                       & People of Europe start to protest, governments begin to introduce first measure \\
    &                    & against the banks, e.g., transaction tax. \\

11 &  Departure                          &  Departure of EU member states from the Eurozone is looming which is  \\
     &                                                &  a threat to some EU member states and a hope of others.\\

12 &  First function of the donor & Some and some politicians representatives of the EU signal that they understand    \\
      &                             & the deplorable situation of those EU countries that are in deep economic trouble.\\

13 &  Hero's reaction                 & People in Europe don't react in a uniform way. Some respond with revival     \\
    &                                                & of national  stereotypes, anger, riots and strikes, others leave their country. \\

14 &  Receipt of a magical agent  & Slowly but steadily rules and regulations are put in place that control the banks.\\
      &                                                    &  \\

15 &  Guidance                          & Guidance and recipes how to overcome the crisis are offered by various parties.  \\
      &                                                &  \\

 \end{tabular}

  \caption{Euro crisis concepts mapped to Propp's narratemes (I)}
\label{Tab:Propp1}
\end{table}

\pagebreak
\begin{table}[hbt]
 \begin{tabular}{ccl}
 No & Narrateme                        & Euro Crisis  \\  \hline
  1 & Absentation                        & national currency, dignity, loss of independence  \\
     &                                               & \\
  2 &  Interdiction                         &  treaty, Maastricht, rules, contract, agreement, BIP, GDP \\
     &                                               &  \\
  3 &  Violation of Interdiction   & budget, deficit, debt, loose spending, squandering, unaffordable \\
     &                                               & \\

4  &  Reconnaissance               & economic, situation   \\
    &                                                & \\

5 &  Delivery                                &  Moodies, Fitch, ratings, S\&\hspace{-0.03cm}P \\
    &                                                & \\
   
6  &  Trickery                                & saving, structural reforms, productivity \\
    &                                                & \\

7  &  Complicity                           &  bailout, IMF, pro austerity, troika\\
    &                                                &   \\

8 &  Villainy                                & austerity, interest rates, lack of growth, saving, unemployment, job losses  \\
    &                                                & \\

9 &  Mediation                          & agreement, summit , talk \\
    &                                                &  \\

10 & Beginning  counteraction & protest, government, transaction tax  \\
    &                    &  \\

11 &  Departure                          &  grexit, leave, threat, hope  \\
     &                                                &   \\

12 &  First function of the donor & compassion, growth, sympathy \\
      &                             & \\

13 &  Hero's reaction                 & anger,  chaos, clashes, collapse, emigration, riot, strike, suicide, turmoil    \\
    &                                                &  \\

14 &  Receipt of a magical agent  & ESF, ESM, euro bonds, fiscal pact, final union, transaction tax,  \\
      &                                                    &  \\

15 &  Guidance                          &   haircut, referendum\\
      &                                                &  \\

 \end{tabular}

  \caption{Euro crisis concepts matching Propp's narratemes (II)}
\label{Tab:Propp2}
\end{table}

\pagebreak
\begin{table}[hbt!]

  \begin{minipage}[t]{140mm}

AAA,
Ackermann,
adapt,
adjustment,
against,
agree,
agreement,
ailing,
alarm,
alarmed, 
alarming,
Albania,
allied,
allies,
ally,
Ansip,
anti+austerity,
appropriate,
Athens,
austere,
austerity,
avert,
averted,
awful,
backward,
bad,
bailout,
bank,
Bank+of+America,
Bank+of+England,		
banking,
bankrupt,
bankruptcy,
banks,
Barclays,
Barroso,
Belgium,
Berlin,
Berlusconi,
Bernanke,
better,
billion,
billions,
BIP,
blackmail,
blame,
bleak,
bonds,
borrowing,
Brazil,
breakthrough,
Britain,
British,
Brussels,
bubble,
budget,
Bundestag,
burden,
burn,
burned,
business,
CAC,
Cameron,
Canada,
capital,
capitulate,
capitulation,
care,
careful,
catastrophe,
catastrophic,
CDS,
CDU,
chaos,
cheat,
cheated,
cheating,
China,
Christofias,
Citigroup,
clash,
clashes,
Coelho,
collapse,
compassion,
compassionate,
competence,
competent,
compromise,
concern,
concerned,
confidence,
confident,
conflict,
contagion,
contagious,
contingency+plan,
contingency+plans,
cooked+the+books,
cooperate,
cooperation,
crash,
Credit+Suisse,
creditors, 			
crisis, 
cruel,
cruelty,
cuts,
Cyprus,
Czech,
danger,
dangerous,
dead,
deal,
death,
debt,
debts,
decline,
declining,
default,
deficit,
Denmark,
depressed,
depressing,
depression,
despair,
destroy,
destroyed,
destruction,
destruction,
Deutsche+Bank,
development,
developments,
die,
died,
difficult,
difficulties,
dignity,
disagree,
disagreement,
disappoint,
disappointed,
disappointment,
disaster,
dismal,
Dollar,
Dollars,
doom,
doomed,
doomsday,
double+dip,
down,
downgrade,
downgraded,
downgrades,
downturn,
Drachma,
Draghi,
drama,
dramas,
dramatic,
dramatically,
drop,
dropped,
ease,
eased,
ECB,
economic,
economical,
economies,
economy,
EFSF,
election+in+France,
election+in+Germany,
election+in+Greece,
election+in+the+US,
emigrate,
emigration,
employed,
employment,
encourage,
encouragement,
end,
end+of+the+eurozone,
endanger,
endangered,
enemies,
enemy,
English,
ESM,
Estonia,
Euro,
euro+zone,
eurobonds,
European+Central+Bank,
eurozone,
exit,
exorbitant,
expensive,
explode,
explosion,
explosive,
fail,
failed+state,
failure,
fair,
fall,
falls,
fatal,
Faymann,
fear,
fears,
Fed,
Federal+Reserve,
fell,
fiasco,
fight,
finance,
financial,
Finland,
fire,
fiscal+cliff,
fiscal+pact, 				
fiscal+union,
Fitch,
forward,
foul,
Franc,
fraud,
French,
fresh,
futile,
gain,
gained,
GDP,
Geithner,
German,
Germany, 
gloom,
gloomy,
gold,
Goldman+Sachs,
good,
government,
Greece,
greed,
greedy,
Greek,
green+shoots,
grexit,
grim,
growth,
haircut,
happy,
harsh,
headway,
health,
healthy,
hedge+fund,
hedge+funds,
help,
helpless,
high,
higher,
Hollande,
honest,
honesty,
honor,
hope,
hope+for+the+eurozone,		
hopes,
horrible,
hostile,
HSBC,
Hungary,
hunger,
hungry,
hurdle,
Iceland,
ill,
IMF,
immigrate,
immigration,
implode,
implosion,
improve,
improved,
improvement,
inappropriate,
incompetence,
incompetent,
India,
industrious,
inflation,
insecure,
instability,
instable,
insult,
insulted,
integrity,
interest,
interests,
interest+rates,		
in+vain,
investment,
investor,
investors,
Ireland,
irresponsible,
ISDA,
Italian,
Italy,
Japan,
Jiabao,
job,
job+losses,
jobs,
JP+Morgan,
Juncker,
Katainen,
Kenny,
Lagarde,
Latvia,
leave,
left+wing,
liar,
lie,
lied,
Lisbon,
Lithuania,
Lloyds,
loathing,
London,
loose+spending,
looting,
lose,
loss,
losses,
lost,
low,
luck,
lucky,
Luxembourg,
Maastricht,
Madrid,
Malta,
market,
markets,
Marshall+plan,
meltdown,
Merkel,
military,
miserable,
misery,
miss,
missed,
misunderstanding,
money,
Monti,
Moodys,				
Morgan+Stanley,
national+currency,
need+help,
negative,
Netherlands,
new+start,
Noda,
Norway,
no+sympathy,
not+happy,
Obama,
odds,
Oil,
optimism,
optimistic,
Orban,
Pahor,
pain,
painful,
Pandora,
Papademos,
Papandreou,
Paris,
peace,
pensions,
people,
pessimism,
pessimistic,
pity,
plummet,
plummeting,
plunder,
Poland,
police,
popular,
Portugal,
positive,
Pound,
powerless,
pressure,
pro+austerity,
problem,
problems,
profit,
progress,
protests,
provocateur,
provocation,
provocative,
provoke,
provoked,
Putin,
quit,
radical,
Rajoy,
ratings,
RBS,
reason,
reasonable,
recession,
reckless,
reconcile,
recover,
recovery,
reduced+salaries,
reduced+salary,
redundancies,
redundant,
referendum,
reform,
reforms,
refusal,
refusals,
refuse,
regression,
Rehn,
Reinfeldt,
reject,
rejection,
relentless,
rescue,
respect,
restricted,
restriction,
right+wing,
riots,
rise,
risk,
risks,
Rome,
Romney,
Rompuy,
row,
Royal+Bank+of+Scotland,
ruin,
ruined,
Rumania,
Rupo,
Rutte,
sad,
sadly,
safe,
Samaras,
Sarkozy,
saving,
savings,
secure,
severe,
shit,
shock,
shortage,
shrank,
shrink,
sick,
Singh,
sink,
sinks,
sleepless,
Slovakia,
Slovenia,
slump,
Societe+General,
solution,
solved,
SP,				
Spain,
Spanish,
SPD,
speculation,
speculative,
spend+thrift,
spending+cuts
Spexit,
spread,
squander,
squandered,
stability,
stable,
stagnate,
stagnation,
stall,
stalled,
steal,
Steinbrueck,
stock,
stock+markets,		
stocks,
stole,
stolen,
storm,
strikes,
strong,
stronger,
struggle,
success,
successful,
suicidal,
suicide,
summit,
summits,
super,
support,
supported,
supporter,
survival,
survive,
survived,
Sweden,
Switzerland,
sympathetic,
sympathy,
Syriza,
talks,
tax,
taxes,
teargas,
terrible,
terrifying,
Thorning-Schmidt,
threat,
threatening,
threats,
tight,
tough,
tragedy,
transaction+tax,
treaty,
treaty+of+Maastricht,
troika,
trouble,
troubles,
Tsipras,
turmoil,
UBS,
UK,
unaffordable,
uncertain,
uncertainty,
unemployed,
unemployment,
unfair,
unhappy,
unions,
unlucky,
unpopular,
unresolvable,
unsafe,
unwilling,
up,
upbeat,
US+economy,
USA,
victim,
victims,
violence,
violent,
war,
wars,
weak,
weaken,
weaker,
weakest,
Weidmann,
wonder,
wonderful,
worried,
worries,
worry,
worse,
worthless,
wreck,
wrecked,
Yen,
Yuan

\end{minipage}
\caption{Total list of concepts used for Twitter analysis (long list)}
\label{Tab:Total_List_of_Concepts}
\end{table}

% ------------------------------------------------------------------- Residual Diagnotics -------------------------------------------------- #

\begin{table}[ht]
\centering
\begin{tabular}{rccccc}
  \hline
\# & R & mean & var & skewness & kurtosis \\ 
  \hline
1 & 310 & -6.406e-20 & 1.917e-05 & 6.777e-09 & 1.51e-09 \\ 
  2 & 320 & -1.228e-19 & 1.946e-05 & 3.011e-09 & 1.512e-09 \\ 
  3 & 330 & -5.385e-22 & 1.835e-05 & 1.781e-09 & 1.363e-09 \\ 
  4 & 340 & -7.64e-20 & 1.98e-05 & 5.837e-09 & 1.498e-09 \\ 
  5 & 350 & 3.008e-20 & 1.987e-05 & -3.045e-09 & 1.61e-09 \\ 
  6 & 360 & 1.136e-20 & 1.956e-05 & 3.405e-11 & 1.494e-09 \\ 
  7 & 370 & 9.425e-20 & 1.902e-05 & -8.709e-10 & 1.465e-09 \\ 
  8 & 380 & -9.465e-21 & 1.873e-05 & 7.19e-09 & 1.474e-09 \\ 
  9 & 390 & -1.456e-21 & 1.865e-05 & 8.431e-09 & 1.547e-09 \\ 
  10 & 400 & 2.582e-20 & 1.838e-05 & 1.035e-08 & 1.434e-09 \\ 
  11 & 410 & 6.3e-20 & 1.821e-05 & 1.537e-08 & 1.377e-09 \\ 
  12 & 420 & -2.25e-20 & 1.906e-05 & 1.694e-08 & 1.605e-09 \\ 
  13 & 430 & 1.892e-20 & 1.881e-05 & 1.638e-08 & 1.623e-09 \\ 
  14 & 440 & -1.047e-19 & 1.91e-05 & 1.729e-08 & 1.706e-09 \\ 
  15 & 450 & 1.847e-20 & 1.927e-05 & 1.722e-08 & 1.728e-09 \\ 
  16 & 460 & -7.407e-20 & 1.977e-05 & 1.823e-08 & 1.753e-09 \\ 
  17 & 470 & -5.053e-20 & 1.872e-05 & 6.603e-09 & 1.68e-09 \\ 
  18 & 480 & 5.564e-20 & 1.795e-05 & -5.051e-09 & 1.46e-09 \\ 
  19 & 490 & -6.49e-20 & 1.812e-05 & -4.492e-09 & 1.52e-09 \\ 
  20 & 500 & -3.676e-20 & 1.715e-05 & -7.909e-09 & 1.536e-09 \\ 
  21 & 510 & -4.177e-20 & 1.67e-05 & -2.201e-08 & 1.558e-09 \\ 
  22 & 520 & -6.05e-21 & 1.775e-05 & -2.509e-08 & 1.587e-09 \\ 
  23 & 530 & 6.537e-20 & 1.787e-05 & -2.676e-08 & 1.743e-09 \\ 
  24 & 540 & 6.069e-20 & 1.935e-05 & -2.753e-08 & 2.144e-09 \\ 
  25 & 550 & 1.025e-19 & 2.06e-05 & -3.184e-08 & 2.171e-09 \\ 
  26 & 560 & 5.722e-20 & 2.025e-05 & -3.251e-08 & 2.348e-09 \\ 
  27 & 570 & 1.792e-20 & 1.997e-05 & -2.541e-08 & 2.313e-09 \\ 
  28 & 580 & -5.395e-20 & 2.191e-05 & -7.81e-08 & 2.17e-09 \\ 
  29 & 590 & -7.422e-20 & 1.986e-05 & -1.006e-07 & 1.999e-09 \\ 
  30 & 600 & 7.623e-21 & 2.531e-05 & -1.191e-07 & 2.831e-09 \\ 
  31 & 610 & 6.899e-20 & 2.567e-05 & -1.342e-07 & 2.808e-09 \\ 
  32 & 620 & -2.711e-20 & 2.697e-05 & -2.683e-07 & 4.165e-09 \\ 
   \hline
\end{tabular}
\caption{First four standardized central residual moments of covariate {\it SP}} 
\label{Moments_SP}
\end{table}

\pagebreak

\begin{table}[ht]
\centering
\begin{tabular}{rccccc}
  \hline
\# & R & mean & var & skewness & kurtosis \\ 
  \hline
1 & 310 & -2.508e-20 & 1.913e-05 & 5.217e-09 & 1.509e-09 \\ 
  2 & 320 & -5.829e-20 & 1.909e-05 & 7.14e-09 & 1.465e-09 \\ 
  3 & 330 & 5.179e-20 & 1.892e-05 & 8.129e-11 & 1.362e-09 \\ 
  4 & 340 & -5.485e-20 & 1.944e-05 & 2.62e-09 & 1.484e-09 \\ 
  5 & 350 & -5.157e-20 & 1.932e-05 & -3.308e-09 & 1.509e-09 \\ 
  6 & 360 & -8.615e-20 & 1.907e-05 & 3.25e-09 & 1.485e-09 \\ 
  7 & 370 & -1.25e-19 & 1.911e-05 & 5.255e-09 & 1.524e-09 \\ 
  8 & 380 & 9.143e-21 & 1.836e-05 & 8.827e-09 & 1.475e-09 \\ 
  9 & 390 & 1.624e-20 & 1.895e-05 & 9.469e-09 & 1.513e-09 \\ 
  10 & 400 & -1.355e-20 & 1.834e-05 & 1.146e-08 & 1.457e-09 \\ 
  11 & 410 & -1.941e-20 & 1.814e-05 & 1.508e-08 & 1.445e-09 \\ 
  12 & 420 & -3.222e-20 & 1.937e-05 & 1.909e-08 & 1.714e-09 \\ 
  13 & 430 & 7.461e-20 & 1.901e-05 & 1.613e-08 & 1.69e-09 \\ 
  14 & 440 & -5.647e-22 & 1.91e-05 & 2.034e-08 & 1.745e-09 \\ 
  15 & 450 & 5.6e-20 & 1.903e-05 & 1.887e-08 & 1.723e-09 \\ 
  16 & 460 & 5.074e-20 & 1.981e-05 & 1.913e-08 & 1.83e-09 \\ 
  17 & 470 & 1.121e-20 & 1.891e-05 & 6.623e-09 & 1.666e-09 \\ 
  18 & 480 & -8.399e-20 & 1.802e-05 & -4.221e-09 & 1.471e-09 \\ 
  19 & 490 & 2.348e-20 & 1.806e-05 & -5.68e-09 & 1.514e-09 \\ 
  20 & 500 & 4.251e-20 & 1.855e-05 & -1.153e-08 & 1.671e-09 \\ 
  21 & 510 & -2.933e-20 & 1.716e-05 & -1.473e-08 & 1.559e-09 \\ 
  22 & 520 & -1.839e-20 & 1.77e-05 & -2.271e-08 & 1.592e-09 \\ 
  23 & 530 & -4.996e-20 & 1.77e-05 & -2.604e-08 & 1.703e-09 \\ 
  24 & 540 & -4.832e-20 & 1.86e-05 & -2.291e-08 & 1.974e-09 \\ 
  25 & 550 & 9.586e-21 & 2.001e-05 & -2.194e-08 & 2.201e-09 \\ 
  26 & 560 & 1.167e-20 & 1.856e-05 & -2.308e-08 & 2.127e-09 \\ 
  27 & 570 & 3.279e-21 & 2.168e-05 & -3.146e-08 & 2.621e-09 \\ 
  28 & 580 & 4.014e-20 & 2.196e-05 & -6.389e-08 & 2.208e-09 \\ 
  29 & 590 & 6.05e-20 & 1.916e-05 & -9.162e-08 & 1.952e-09 \\ 
  30 & 600 & 2.446e-20 & 2.299e-05 & -1.208e-07 & 2.515e-09 \\ 
  31 & 610 & -2.07e-19 & 2.401e-05 & -1.323e-07 & 2.446e-09 \\ 
  32 & 620 & 5.421e-20 & 2.649e-05 & -2.409e-07 & 3.787e-09 \\ 
   \hline
\end{tabular}
\caption{First four standardized central residual moments of covariate {\it risk}} 
\label{Moments_risk}
\end{table}

\begin{table}[ht]
\centering
\begin{tabular}{p{1cm}   c   c   c   c   c   c   c   c   c   c   c   }

  \hline
out & $R$ & $P$ & $\bar{x}$ res & LB/raw & LB/adj & KP/raw & KP/adj & WH/raw & WH/adj \\ 
  \hline
 + & 310 & 322 & 0.000 & 0.004 & 0.122 & 0.100 & 1.000 & 0.487 & 1.000 \\ 
- & 320 & 312 & 0.000  & 0.004 & 0.139 & 0.100 & 1.000 & 0.876 & 1.000 \\ 
 + & 330 & 302 & 0.000 & 0.004 & 0.123 & 0.100 & 1.000 & 0.820 & 1.000 \\ 
 + & 340 & 292 & 0.000 & 0.004 & 0.122 & 0.100 & 1.000 & 0.631 & 1.000 \\ 
 + & 350 & 282 & 0.000 & 0.001 & 0.047 & 0.100 & 1.000 & 0.866 & 1.000 \\ 
 - & 360 & 272 & 0.000  & 0.003 & 0.112 & 0.100 & 1.000 & 0.991 & 1.000 \\ 
 + & 370 & 262 & 0.000 & 0.019 & 0.609 & 0.100 & 1.000 & 0.746 & 1.000 \\ 
 + & 380 & 252 & 0.000 & 0.000 & 0.011 & 0.100 & 1.000 & 0.817 & 1.000 \\ 
 + & 390 & 242 & 0.000 & 0.002 & 0.055 & 0.100 & 1.000 & 0.975 & 1.000 \\ 
 + & 400 & 232 & 0.000 & 0.006 & 0.187 & 0.100 & 1.000 & 0.650 & 1.000 \\ 
+ & 410 & 222 & 0.000  & 0.008 & 0.266 & 0.100 & 1.000 & 0.567 & 1.000 \\ 
 + & 420 & 212 & 0.000 & 0.020 & 0.648 & 0.100 & 1.000 & 0.759 & 1.000 \\ 
 - & 430 & 202 & 0.000  & 0.018 & 0.585 & 0.100 & 1.000 & 0.723 & 1.000 \\ 
 - & 440 & 192 & 0.001  & 0.027 & 0.876 & 0.100 & 1.000 & 0.815 & 1.000 \\ 
 - & 450 & 182 & 0.000  & 0.015 & 0.494 & 0.100 & 1.000 & 0.848 & 1.000 \\ 
 - & 460 & 172 & 0.000  & 0.037 & 1.000 & 0.100 & 1.000 & 0.706 & 1.000 \\ 
 - & 470 & 162 & 0.000  & 0.077 & 1.000 & 0.100 & 1.000 & 0.885 & 1.000 \\ 
 - & 480 & 152 & 0.000  & 0.097 & 1.000 & 0.100 & 1.000 & 0.877 & 1.000 \\ 
 + & 490 & 142 & 0.000 & 0.046 & 1.000 & 0.100 & 1.000 & 0.857 & 1.000 \\ 
 + & 500 & 132 & 0.000 & 0.072 & 1.000 & 0.100 & 1.000 & 0.790 & 1.000 \\ 
 + & 510 & 122 & 0.000 & 0.102 & 1.000 & 0.100 & 1.000 & 0.735 & 1.000 \\ 
 - & 520 & 112 & 0.000  & 0.231 & 1.000 & 0.100 & 1.000 & 0.670 & 1.000 \\ 
 + & 530 & 102 & 0.001 & 0.131 & 1.000 & 0.100 & 1.000 & 0.732 & 1.000 \\ 
 - & 540 & 92 & 0.000 & 0.224 & 1.000 & 0.100 & 1.000 & 0.501 & 1.000 \\ 
 + & 550 & 82 & 0.000  & 0.156 & 1.000 & 0.100 & 1.000 & 0.741 & 1.000 \\ 
 + & 560 & 72 & 0.000 & 0.409 & 1.000 & 0.100 & 1.000 & 0.637 & 1.000 \\ 
 + & 570 & 62 & 0.000  & 0.371 & 1.000 & 0.100 & 1.000 & 0.608 & 1.000 \\ 
 + & 580 & 52 & 0.000  & 0.251 & 1.000 & 0.100 & 1.000 & 0.946 & 1.000 \\ 
 - & 590 & 42 & 0.000  & 0.562 & 1.000 & 0.090 & 1.000 & 0.993 & 1.000 \\ 
 - & 600 & 32 & 0.000  & 0.387 & 1.000 & 0.058 & 1.000 & 0.923 & 1.000 \\ 
 - & 610 & 22 & -0.001 & 0.735 & 1.000 & 0.100 & 1.000 & 0.611 & 1.000 \\ 
 + & 620 & 12 & -0.001 & 0.554 & 1.000 & 0.100 & 1.000 & 0.203 & 1.000 \\ 
\hline
\multicolumn{10}{l}{%
  \begin{minipage}{12.5cm}%
    \small $out$ indicates whether the RW model has been outperformed (+) or not (-), 
    LB; Ljung-Box test for autocorrelation,
    KP: KPSS test for trend stationarity, 
    WH: White neural network test for nonlinearity, 
    raw: raw $p$-values, 
    adj: Bonferroni adjusted $p$-values.%
  \end{minipage}%
}\\
\end{tabular}
\caption{Residual diagnostics of the Tweet count time series of {\it SP}} 
\label{Residual_Diagnostics_SP}
\end{table}

\begin{table}[ht]
\centering
\begin{tabular}{cccccccccc}
  \hline
out & $R$ & $P$ & $\bar{x}$ res & LB/raw & LB/adj & KP/raw & KP/adj & WH/raw & WH/adj \\ 
  \hline
+ & 310 & 322 & 0.000 & 0.004 & 0.122 & 0.100 & 1.000 & 0.529 & 1.000 \\ 
+ & 320 & 312 & 0.000  & 0.004 & 0.139 & 0.100 & 1.000 & 0.834 & 1.000 \\ 
+ & 330 & 302 & 0.000  & 0.004 & 0.123 & 0.100 & 1.000 & 0.690 & 1.000 \\ 
+ & 340 & 292 & 0.000  & 0.004 & 0.122 & 0.100 & 1.000 & 0.619 & 1.000 \\ 
+ & 350 & 282 & 0.000  & 0.001 & 0.047 & 0.100 & 1.000 & 0.826 & 1.000 \\ 
+ & 360 & 272 & 0.000  & 0.003 & 0.112 & 0.100 & 1.000 & 0.952 & 1.000 \\ 
+ & 370 & 262 & 0.000  & 0.019 & 0.609 & 0.100 & 1.000 & 0.718 & 1.000 \\ 
+ & 380 & 252 & 0.000  & 0.000 & 0.011 & 0.100 & 1.000 & 0.871 & 1.000 \\ 
+ & 390 & 242 & 0.000  & 0.002 & 0.055 & 0.100 & 1.000 & 0.830 & 1.000 \\ 
+ & 400 & 232 & 0.000  & 0.006 & 0.187 & 0.100 & 1.000 & 0.638 & 1.000 \\ 
+ & 410 & 222 & 0.000 & 0.008 & 0.266 & 0.100 & 1.000 & 0.602 & 1.000 \\ 
- & 420 & 212 & 0.000 & 0.020 & 0.648 & 0.100 & 1.000 & 0.884 & 1.000 \\ 
- & 430 & 202 & 0.000  & 0.018 & 0.585 & 0.100 & 1.000 & 0.928 & 1.000 \\ 
- & 440 & 192 & 0.001  & 0.027 & 0.876 & 0.100 & 1.000 & 0.785 & 1.000 \\ 
- & 450 & 182 & 0.000  & 0.015 & 0.494 & 0.100 & 1.000 & 0.820 & 1.000 \\ 
- & 460 & 172 & 0.000 & 0.037 & 1.000 & 0.100 & 1.000 & 0.856 & 1.000 \\ 
- & 470 & 162 & 0.000 & 0.077 & 1.000 & 0.100 & 1.000 & 0.863 & 1.000 \\ 
- & 480 & 152 & 0.000  & 0.097 & 1.000 & 0.100 & 1.000 & 0.858 & 1.000 \\ 
+ & 490 & 142 & 0.000 & 0.046 & 1.000 & 0.100 & 1.000 & 0.926 & 1.000 \\ 
- & 500 & 132 & 0.000  & 0.072 & 1.000 & 0.100 & 1.000 & 0.942 & 1.000 \\ 
- & 510 & 122 & 0.000  & 0.102 & 1.000 & 0.100 & 1.000 & 0.772 & 1.000 \\ 
- & 520 & 112 & 0.000  & 0.231 & 1.000 & 0.100 & 1.000 & 0.388 & 1.000 \\ 
+ & 530 & 102 & 0.001 & 0.131 & 1.000 & 0.100 & 1.000 & 0.543 & 1.000 \\ 
+ & 540 & 92 & 0.000  & 0.224 & 1.000 & 0.100 & 1.000 & 0.659 & 1.000 \\ 
+ & 550 & 82 & 0.000  & 0.156 & 1.000 & 0.100 & 1.000 & 0.722 & 1.000 \\ 
+ & 560 & 72 & 0.000  & 0.409 & 1.000 & 0.100 & 1.000 & 0.554 & 1.000 \\ 
- & 570 & 62 & 0.000  & 0.371 & 1.000 & 0.100 & 1.000 & 0.841 & 1.000 \\ 
+ & 580 & 52 & 0.000 & 0.251 & 1.000 & 0.100 & 1.000 & 0.900 & 1.000 \\ 
+ & 590 & 42 & 0.000 & 0.562 & 1.000 & 0.090 & 1.000 & 0.577 & 1.000 \\ 
- & 600 & 32 & 0.000  & 0.387 & 1.000 & 0.058 & 1.000 & 0.869 & 1.000 \\ 
- & 610 & 22 & -0.001 & 0.735 & 1.000 & 0.100 & 1.000 & 0.597 & 1.000 \\ 
+ & 620 & 12 & -0.001& 0.554 & 1.000 & 0.100 & 1.000 & 0.201 & 1.000 \\ 
\hline
\multicolumn{10}{l}{%
  \begin{minipage}{12.5cm}%
    \small $out$ indicates whether the RW model has been outperformed (+) or not (-), 
    LB; Ljung-Box Test for autocorrelation,
    KP: KPSS test for trend stationarity, 
    WH: White neural network test for nonlinearity, 
    raw: raw $p$-values, 
    adj: Bonferroni adjusted $p$-values.%
  \end{minipage}%
}\\
\end{tabular}
\caption{Residual diagnostics of the Tweet count time series of {\it risk}} 
\label{Residual_Diagnostics_risk}
\end{table}

\end{document}